\documentclass[reprint,10pt,twocolumn,superscriptaddress,nofootinbib]{revtex4-1}

\usepackage[utf8]{inputenc}
\usepackage{graphicx}
\usepackage{bm}
\usepackage{caption}

\usepackage{amsmath,amssymb,amsbsy}
\usepackage{bbm,mathrsfs,yfonts}
\usepackage{float}
\usepackage{silence}
\usepackage{mathtools}

\renewcommand{\vec}[1]{{\boldsymbol{#1}}}

\begin{document}

% \preprint{}
%\begin{frontmatter}

\title{Lattice Boltzmann model for collisionless electrostatic drift wave turbulence obeying Charney-Hasegawa-Mima dynamics}

\author{M. Held }
\email{markus.held@uibk.ac.at}
\author{A. Kendl}
\email{alexander.kendl@uibk.ac.at}
\affiliation{Institute for Ion Physics and Applied Physics, Association Euratom-\"OAW,
 University of Innsbruck, A-6020 Innsbruck, Austria}
%\fntext[fn1]{Phone number: +43 (0) 512 507 52727}
%\fntext[fn2]{Phone number: +43 (0) 512 507 52720}

%\date{\today}

\begin{abstract}
A lattice Boltzmann method (LBM) approach to the Charney-Hasegawa-Mima (CHM)
model for adiabatic drift wave turbulence in magnetised
plasmas, is implemented.
The CHM-LBM model contains a barotropic equation of state for the
potential, a force term including a cross-product analogous to the Coriolis
force in quasigeostrophic models, and a density gradient source term.
Expansion of the resulting lattice Boltzmann model equations leads to cold-ion
fluid continuity and momentum equations, which resemble CHM dynamics under
drift ordering. 
The resulting numerical solutions of standard test cases (monopole propagation,
stable drift modes and decaying turbulence) are compared
to results obtained by a conventional finite difference scheme that 
directly discretizes the CHM equation. 
The LB scheme resembles characteristic CHM dynamics apart from an additional shear in the density gradient direction. The occuring shear
reduces with the drift ratio and is ascribed to the compressible limit of the underlying LBM.
\end{abstract}
\maketitle

\section{\label{sec:intro}Introduction}

The lattice Boltzmann method (LBM) has been established as a promising tool
for computations in fluid dynamics, including turbulence, reactive and complex
flows. The LB method to model fluid partial differential equations in the
framework of a reduced discrete kinetic theory has also been applied
to plasma physics. 
Problems like magnetohydrodynamic turbulence 
(treated for example in refs.~\cite{chen91,succi91,martinez94,vahala00,schaffenberger02,dellar02pre,
dellar03,breyiannis04,macnab06,vahala08,pattison08,dellar09,chatterjee10,dellar11}), 
magnetic reconnection \cite{mendoza08,dellar02jcp,dellar13}, and a first approach to electrostatic turbulence \cite{fogaccia96}
have been addressed in this framework. 

The Charney-Hasegawa-Mima (CHM) equation serves as a basic prototypical two-dimensional
one-field model for collisionless electrostatic drift wave turbulence in magnetised plasmas
with cold ions and isothermal electrons with an adiabatic response.
Drift wave turbulence taps free energy from the background plasma pressure gradient to
drive advective nonlinear motion  of pressure disturbances by the $E \times B$ drift velocity perpendicular
to the magnetic field $\vec{B}$. Parallel dynamics are captured by the electron currents
which are balancing the pressure deviations electrostatically with an adiabatic response 
along the magnetic field. The spatial scale is highly anisotropic permitting us to decouple the parallel dynamics from
the perpendicular drift plane motion obeying the two-dimensional (normalized) CHM equation \cite{hasegawa77,charney48}
\begin{equation} \label{chm}
 \Big(1- \nabla^2 \Big) \frac{ \partial \delta\phi}{\partial t} + 
\frac{ \partial \delta \phi}{\partial y} - \Big\{ \delta \phi, \nabla^2 \delta
\phi \Big\} = 0
\end{equation}
where the advective nonlinearity is expressed by a Poisson bracket 
$\{A,B\} = \partial_x A \; \partial_y B - \partial_y A \; \partial_x B$.
The equation is normalized according to  $\vec{x} \leftarrow  \vec{x} / \rho_s $ and 
$ t \leftarrow \kappa_n \omega_{ci} t$ for the length and time scales 
and fluctuations $\delta \phi \leftarrow \kappa_n^{-1}  (e \phi /T_e)$ for the
electrostatic potential $\phi$.
These scales represent the dominant contributions to turbulent transport in magnetised plasmas, where the drift frequency
\(\omega \sim (\rho_s / L_n) \hspace{1 mm} \omega_{ci}\) appears
to be lower in magnitude then the ion gyro frequency \(\omega_{ci} = c_s / \rho_s\) describing the gyro-motion of ions around the magnetic field lines. The magnitude
is specified by the ratio of the drift scale \(\rho_s = \sqrt{m_i T_e} / eB\) (corresponding to a gyro radius of ions of mass
\(m_i$ at electron temperature $T_e\)) to the gradient length \(L_n = |\partial_x \ln n_0(x)|^{-1}\) of the static background density $n_0(x)$ and is typically defined by the drift ratio
\(\kappa_n = \rho_s / L_n\ll 1\). The sound speed $c_s = \sqrt{T_e / m_i}$ is given in terms of the electron temperature and ion mass. Finite ion temperature (\(T_i>0\)) effects arise
when the ion gyro-radius \(\rho_i = \sqrt{m_i T_i} / eB\) approaches typical fluctuation scales and are beyond the scope of the model. More detailed gyrokinetic or gyrofluid models put
emphasize on accurate averaging procedures over gyro-motion and modifications to the the polarization equation \cite{kendl08,scott07,krommes12}.
\\
The CHM equation can be either obtained from a gyrokinetic model, or from the
continuity and momentum equations for a cold uniformally magnetised ion fluid ($T_i \ll T_e$) with
adiabatic electron response and a negative background density gradient in
x-direction, $n_i = n_e = n_0 (x) \exp[e \phi/kT_e]$. 
The normalized ion continuity and momentum equations can be expressed in
terms of the potential instead of density \cite{horton94} as
\begin{eqnarray}
 \kappa_n \frac{d}{dt} \delta \phi + {\vec\nabla } \cdot {\vec{u}} &=&
 \kappa_n  \hspace{1 mm} {\vec{u} } \cdot  {\vec\nabla }
 {x} \label{ioncont} \\
  \kappa_n \frac{d}{dt}    {\vec{u}}  +  \vec{e}_z \times
  {\vec{u}} &=& -  {\vec\nabla}\delta \phi \label{ionmom}
\end{eqnarray}
where $d/dt = \partial_t + \vec{u} \cdot {\vec\nabla }$ is the advective derivative.
Expanding $\delta \phi$ and $\vec{u}$ in an asymptotic series
with the drift ratio $\kappa_n \ll 1$ as small expansion parameter and
accordant ordering \cite{horton94} yields the CHM eq.~(\ref{chm}). 
Replacing the drift ratio \(\kappa_n\) with the Rossby number \(\mathit{Ro}\) and identifying the electrostatic potential fluctuations 
with the dimensionless surface height reveals the isomorphism to the quasi-geostrophic single layer shallow water equations in the \(\beta\)-plane approximation.
By replacing the density gradient with a bottom topography or a spatially varying Coriolis frequency the CHM equation is resembled in the limit
of a small Rossby number \(\mathit{Ro} \ll 1\). Advances with the Lattice Boltzmann method to the shallow water equations have been made by Zhong et al. 
\cite{zhong05,zhong06a,zhong06b} and Dellar \cite{dellar00}.

\section{\label{sec:lbm}Lattice Boltzmann Model}

\subsection{Boltzmann equation}
Starting point for the lattice discretization is the Boltzmann equation 
for the kinetic distribution function $f (\vec{x},\vec{\xi},t)$
with a Bhatnagar-Gross-Krook (BGK) collision operator $C = - (f - f^{eq})/\tau_c$,
which expresses the relaxation  to a local Maxwellian for a time constant $\tau_c$. 
Applying the diffusive scaling \(t\rightarrow t/\epsilon^2\) and \(\vec{x} \rightarrow \vec{x}/\epsilon \) on the Boltzmann equation
results in its dimensionless form \cite{bhatnagar54}
\begin{equation} 
  {\frac{\partial}{\partial t}} f + \frac{1}{\epsilon}{\vec{\xi}} \cdot {\vec\nabla} f 
   = \frac{1}{\epsilon^2} \left[ A(f- f^{eq}) + F \right],
\label{kineticeq}
\end{equation}
where source and force terms are included in a forcing function as 
$ F(\vec{x},\vec{\xi},t) = - \vec{a} \cdot \vec{\nabla_{\vec{\xi}}}
f(\vec{x},\vec{\xi},t) + {s}(\vec{x},\vec{\xi},t)$ and the single time collision operator is defined by
\( A = - 1/\left(\epsilon\tau \right)  \)

The Knudsen number $\epsilon = \lambda_{m} / L_0$ and the
non-dimensional relaxation time  $\tau = \tau_c /t_c$ are here
defined in relation to characteristic drift scale $L_0=\rho_s$ and to the 
collision time $t_c = \lambda_{m} / U_0$ with mean free path length $\lambda_{m} = e_m \tau_c $
and characteristic (drift) velocity $U_0=\kappa_n c_s$. The dimensionless
relaxation time ${\tau} = U_0 / e_m $ relates the flow
velocity to the (lattice) molecular velocity $e_m $ whereas the Mach number \(\mathit{Ma} = U_0/c_s \) is identified with the drift parameter \(\kappa_n \) .

The dynamics in the fluid limit, given by eqs.~(\ref{ioncont}) and (\ref{ionmom}),
can be consistently described with the kinetic eq.~(\ref{kineticeq}) assuming
a local Maxwellian equilibrium distribution function of the form \cite{dellar02pre}  
\begin{equation}
 f^{eq} =  \frac{\phi}{ ( {2 \pi \Theta} )^{D/2}} 
\exp{\left[ - \frac{ (\vec{\xi}-\vec{u})^2}{2 \Theta} \right]}.
\end{equation}

The squared dimensionless barotropic speed of sound 
${\Theta} =  \phi / (2 \kappa_n^2)$ 
results from the barotropic pressure term ${P} = \phi^2 / (2 \kappa_n^2)$ 
appearing on the macroscopic level as in eq.~(\ref{ionmom}). 
The isothermal squared speed of sound is defined by ${\theta} =
1/\kappa_n^2$. 

Macroscopic quantities are defined by taking velocity moments over the distribution function
\begin{eqnarray}
 \phi &=& \int f d\vec\xi \\
 \phi {\vec{u}} &=& \int \vec{\xi} f d\vec{\xi} \\
 {\vec{\Pi}}^{(0)} &=& \int \vec{\xi}\vec{\xi}  f^{eq} d\vec{\xi} =
  {P} \vec{I}  +\phi {\vec{u}}{\vec{u}}
\end{eqnarray}
and over the forcing function
\begin{eqnarray}
  \int F d\vec{\xi} &=&  \phi s\\
 \int \vec{\xi} F d \vec{\xi} &=& \phi \vec{a}\\
 \int \vec{\xi} \vec{\xi}  F d \vec{\xi} &=& \phi
 (\vec{a}\vec{u}+\vec{u} \vec{a}) + \phi \frac{d P}{d \phi} s \vec{I}.
\end{eqnarray}

\subsection{Lattice Boltzmann equation}

The discretization of the continuum velocity space to 9 directions in 
2 dimensions (D2Q9) casts the set of velocities to 
$ \{{\vec\xi}_0,{\vec\xi}_1,...,{\vec\xi}_8 \}$
and the distribution function and forcing function to  
$f(\vec{x},\vec{\xi}_i,t)/w(\vec{\xi}_i) = f_i(\vec{x},t)/w_i$ and 
$F(\vec{x},\vec{\xi}_i,t)/w(\vec{\xi}_i) = F_i(\vec{x},t)/w_i$ for $i \in (0,...,8)$  
with the continous weight function \(w(\vec{\xi}_i)\) of the Gauss-Hermite  quadrature formula.
The lattice velocities in D2Q9 geometry are
\begin{eqnarray}
{\vec\xi}_0 &=& \left(0,0\right)\\
{\vec\xi}_i &=& \sqrt{3} \; \xi_m \left(\cos \alpha_i, \sin \alpha_i \right) \quad \mbox{for} \quad i=1,2,3,4\\
{\vec\xi}_i &=& \sqrt{6} \; \xi_m \left( \cos \beta_i, \sin \beta_i  \right) \quad \mbox{for} \quad i=5,6,7,8
\end{eqnarray} 
where $\alpha_i = (i-1) \pi/2$ and $\beta_i = \alpha_i + \pi/4$.
The lattice speed of sound is defined by 
$\xi_m = (1/\sqrt{3}) (\delta x / \delta t)$
with lattice grid size $\delta x = L_B /N_x$ and time step $\delta t$.
The numerical box size $L_B$ with $N_x$ grid points per space dimension crucially determines
whether the CHM is resolved in the drift wave limit.

The choice of the correct equilibrium distribution function for a LBM
depends mainly on the equation of state  and the lattice geometry. 
The equilibrium distribution function for a barotropic equation of state acting on a D2Q9 lattice has been determined by
Dellar \cite{dellar02pre} who showed that an augmentation of the hydrodynamic
equilibrium distribution function by ghost modes is leading to a stable scheme if the ghost variables are properly set.

The equilibrium distribution function from ref.~\cite{dellar02pre} equals one previously derived from an
ansatz Method in ref.~\cite{salmon99}:
\begin{eqnarray} 
\label{edfalg}
 f_0^{eq} &=& w_0 \phi \left[ \frac{9}{4} - \frac{5}{4} \frac{{P}(\phi)}{\phi
     {\theta}}  -  \frac{{\vec{u}}^2}{ 2{\theta}}\right] \nonumber\\ 
 f_i^{eq} &=& w_i \phi \left[\frac{{P}(\phi)}{\phi {\theta}}  + \frac{
     {\vec\xi}_i\cdot {\vec{u}}}{{\theta}}  + \frac{ ({\vec\xi}_i \cdot
     {\vec{u}})^2}{2 {\theta}^2}   -\frac{{\vec{u}}^2}{ 2 {\theta}} \right].
\end{eqnarray}

Taking the discrete velocity moments 
\begin{align}
\sum\limits_{i=0}^8  f_i^{eq}  =& \phi  \label{firstmom} \\
\sum\limits_{i=0}^8 f_i^{eq} {\vec\xi}_{i} =& \phi {\vec{u}} \label{secondmom} \\
\sum\limits_{i=0}^8 f_i^{eq} {\vec\xi}_{i } {\vec\xi}_{i}  =& \phi
           {\vec{u}}{\vec{u}} + {P} \vec{I} = {\vec{\Pi}}^{(0)} \label{thirdmom} \\ 
\sum\limits_{i=0}^8 f_i^{eq} {\xi}_{i \alpha} {\xi}_{i \beta} {\xi}_{i \gamma}
=&  {\theta} \phi ({u}_{\alpha} \delta_{\beta\gamma}  + {u}_{\beta}
\delta_{\gamma\alpha}  +  {u}_\gamma \delta_{\alpha \beta}) =  {\Lambda}^{(0)}_{\alpha \beta \gamma} \label{fourthmom} 
\end{align}
reveals the deviation to continous kinetic theory where the third velocity moment reads 
\( \Lambda_{\alpha \beta \gamma}  = {\Theta} \phi ({u}_{\alpha} \delta_{\beta\gamma}  + {u}_{\beta}
\delta_{\gamma\alpha}  +  {u}_\gamma \delta_{\alpha \beta}) +\phi u_\alpha u_\beta u_\gamma \). Hence in the
\(\Lambda_{\alpha \beta \gamma}^{(0)} \) of the D2Q9 lattice model the 
$\mathcal{O}(\epsilon^3)$ triple term is missing and the dimensionless
squared isothermal sound speed  ${\theta}$ appears instead of the squared barotropic
sound speed ${\Theta}$.
These differences are further discussed in ~\ref{app:aa}.

The velocity moments over the discrete form of the forcing function 
determine the force and source terms in the macroscopic equations.
The desired fluid system exhibits a velocity dependent force term 
\(\kappa_n^{-1} \vec{e}_z \times \vec{u}\) containing a cross product and an
additional velocity dependent density gradient source term 
\(\kappa_n \vec{u}\cdot \vec{\nabla}{x}\). The force term is mathematically identical to
the Coriolis force, which was already treated by two- and three-dimensional lattice Boltzmann algorithms
\cite{yu05,salmon99b,dellar13cma,dellar02pre,zhong05,tsutahara01}.

The forcing function proposed in the following resembles the first three velocity moments at
the continuum kinetic level, and hence incorporates the barotropic equation of state
appearing in the second second velocity moment
\begin{eqnarray}
 \sum\limits_{i=0}^8 F_i &=& \phi s, \\
 \sum\limits_{i=0}^8 F_i {\vec\xi}_{i} &=&\phi {\vec{a}} \\
 \sum\limits_{i=0}^8 F_i {\vec\xi}_{i } {\vec \xi}_{i}&=&  \phi \left[{\vec{a}} {\vec{u}}
          + {\vec{u}} {\vec{a}}\right]+ \phi \frac{d P}{d \phi} s \vec{I},
\end{eqnarray}
The forcing function is derived in an analogous manner to the equilibrium distribution function by considering the ghost
variables (see ref. \cite{dellar02pre} for details) and generalizes the forcing function of
Luo \cite{luo98} for complex fluids with a barotropic equation of state. 
\begin{align}
F_i =&   w_i \phi \bigg\{  
 \left[ 1 + \left( {\theta} - \frac{d{P}}{d\phi} \right)
  \left( \frac{4 + g_i}{4 {\theta}}-\frac{{\vec\xi}^2}{2 {\theta}^2} \right)
 \right]{s}  
  \nonumber \\&+ \left[ \frac{ ({\vec\xi}_i - {\vec{u}}) }{{\theta}} + 
\frac{ ( {\vec\xi}_i\cdot {\vec{u}} ) {\vec\xi}_i}{{\theta}^2} \right] 
\cdot {\vec{a}} \bigg\}
\label{forcingfunctionalg}
\end{align}
The normalized CHM  source and force terms are
\begin{eqnarray}
 s &=& \kappa_n \vec{u} \cdot \vec{e}_x,  \label{dimlesslbmsource} \\
 {\vec{a}} &=& \frac{1}{\kappa_n} \left(\vec{u}\times \vec{e}_z \right) + \vec{u} {s}. 
\label{dimlesslbmforce}
\end{eqnarray}
The additional contribution of $\vec{u} {s}$ in the force
term will cancel a spurious term in the macroscopic momentum equation, 
which is detailed in the asymptotic analysis in ~\ref{app:aa}.

\subsection{CHM-LBM time integration}
The diffusively scaled discrete Boltzmann PDE follows from eq. (\ref{kineticeq})
\begin{align}
\label{discreteBoltzscaled}
  \frac{\partial f_i}{\partial t} + \frac{1}{\epsilon}\hspace{1 mm} \vec{\xi}_i \cdot \vec{\nabla} f_i &=
  \frac{1}{\epsilon^2}\left[A (f_i - f_i^{eq}) + F_i\right],
\end{align}
Writing the left hand side as the total derivative 
\(\frac{d}{d s} f_i(\vec{x}+\vec{\xi}_i s/\epsilon, t +s)\) and integrating both sides of eq. (\ref{discreteBoltzscaled}) 
from \(s=0\) to \(s=\delta t = \epsilon^2 \) yields \cite{he98,dellar13cma}
\begin{equation} 
 f_{i} ({\vec{x}}',t') - f_{i} ({\vec{x}},t)  =  \frac{1}{\epsilon^2} \int_0^{\delta t} {h}_i(\vec{x}+\vec{\xi}_i s/\epsilon, t +s) ds .
\label{intkineticeq}
\end{equation}
with the substitution \(h_i = A (f_i - f_i^{eq}) + F_i\) for the right hand side of eq. (\ref{discreteBoltzscaled}). 
The implicit lattice Boltzmann equation is now obtained by approximating the integral by a second order accurate numerical
quadrature scheme. This is ensured by the trapezoidal rule 
\begin{align}
 \frac{1}{\epsilon^2} \int_0^{\delta t} {h}_i(\vec{x}+\vec{\xi}_i s/\epsilon, t +s) ds =&
\frac{1}{2} \left[{h}_i({\vec{x}}',t')  +{h}_i({\vec{x}} ,t) \right] \nonumber \\&+ \mathcal{O}(\epsilon^4)
\end{align}
with the substitution \(({\vec{x}}',t') = (\vec{x}+\vec{\xi}_i \epsilon ,t+\epsilon^2) \). Writing out the integral yields
the implicit form of the lattice Boltzmann equation
\begin{align}
\label{implicitLBEscaled}
  f_{i} ({\vec{x}}',t') - f_{i} ({\vec{x}},t)  =& 
  \frac{1}{2} \big\{A \big[ {f}_i({\vec{x}}',t')  +  {f}_i({\vec{x}} ,t) -
		    f_i^{eq}({\vec{x}}',t') \nonumber \\&- f_i^{eq}({\vec{x}} ,t) \big]
		   +\left[{F}_i({\vec{x}}',t')  +  {F}_i({\vec{x}} ,t) \right]\big\}.
\end{align}
To work around this implicit equation the distribution function is transformed
to $f \rightarrow \bar f$ as
\begin{align}
 \bar f_{i}({\vec{x}},t) = f_{i} ({\vec{x}},t) - 
  \frac{A}{2 }\left[ f_{i} ({\vec{x}},t) - f_{i}^{eq} ({\vec{x}},t) \right]-
  \frac{1}{2} F_{i} ({\vec{x}},t) 
\end{align}

Introducing now the \(\bar A = A/\left[1-1/(2A)\right]\) and \(\bar \lambda =  \bar A / A\)
and applying the transformation $f \rightarrow \bar f$ 
to eq. (\ref{implicitLBEscaled}) resembles the usual form of the explicit LB algorithm
\begin{align}
\label{explicitscaledLBE}
 \bar f_{i}({\vec{x}}',t') -  \bar f_{i} ({\vec{x}},t) =
\bar A  \left[ \bar f_{i}
({\vec{x}},t) - f_{i}^{eq} ({\vec{x}},t) \right] +\bar \lambda F_{i}  ({\vec{x}},t),
\end{align}
which is the starting point of the asymptotic analysis presented in \ref{app:aa}.
However, this transformation leads to implicit expressions of the velocity moments
over the distribution function as
\begin{align}
\label{zerothvelocitymomentalg} 
 \phi =& \sum\limits_{i=0}^N  \bar f_i  + \frac{1}{2}
 \phi {s} \\ 
\label{firstvelocitymomentalg}
 \phi {\vec{u}} =& \sum\limits_{i=0}^N {\vec\xi}_i  \bar f_i + \frac{1}{2} \phi {\vec{a}}  \\ 
 \label{secondvelocitymomentalg}
\left(1-\frac{A}{2}\right) {\vec{\Pi}} =& \sum\limits_{i=0}^N \bar f_i {\vec\xi}_i {\vec\xi}_i 
              - \frac{A}{2 } {\vec{\Pi}}^{(0)} +\frac{1}{2} \phi \left({\vec{a}} {\vec{u}}
              +  {\vec{u}}{\vec{a}}\right) \nonumber \\&+ \frac{1}{2}
              \frac{d {P}}{d { \phi}} \phi  {s} 
\end{align}
Depending on the exact form of the force and source terms these equations may
not have an analytical solution which is at the same time computationally
efficient, and one has to apply Newton's method to this problem. 
In this particular case the relevant equations
\begin{eqnarray}
 \phi &=& \bar\phi+\frac{1}{2} \frac{1}{L_n}  \phi \left( {\vec{u}} \cdot \vec{e}_x\right)\\
 \label{velshift_dwe}
 {\vec{u}} &=& \frac{\bar\phi}{\phi}\bar{\vec{u}} 
  - \frac{1}{2}\omega_{ci} (\vec{e}_z \times {\vec{u}}) 
  +  \frac{1}{2}\frac{1}{L_n}  {\vec{u}
  }\left( {\vec{u}} \cdot \vec{e}_x\right)
\end{eqnarray}
with
\begin{eqnarray}
 \bar\phi &=& \sum_{i=0}^N  \bar f_i \quad\quad \mbox{and} \quad\quad
\bar\phi\bar{\vec{u}}= \sum_{i=0}^N {\vec\xi}_i  \bar f_i \label{phi_u_bar}
\end{eqnarray}
could not be solved trivially. An approximation which stays within the
scope of the model has to be made at this point.  
It is justified to drop the third term on the right hand side of the relation
for the velocity shift eq.~(\ref{velshift_dwe}). By taking the cross product of the approximated expression we obtain
\(\vec{e_z} \times \vec{u} =\left[1-\frac{1}{2} \frac{1}{L_n} \left( {\vec{u}} \cdot \vec{e}_x\right) \right] 
\vec{e_z} \times \vec{\bar{u}}+ \frac{1}{2}  \vec{u}\omega_{ci}  \). This simplifies the
equations for the shifts to: 
\begin{eqnarray} 
 \phi &\gets& \frac{ \bar\phi}{ \displaystyle 1 - \frac{1}{2}
 \frac{1}{L_n}{\vec{u}}\cdot \vec{e}_x }
 \label{approxzerothvelocitymoment} \\ 
 {\vec{u}} &\gets& \frac{\displaystyle \bar{\vec{u}} - \frac{1}{2}
{\omega}_{ci} \left(\vec{e}_z \times \bar{\vec{u}} \right) }{
   \displaystyle 1+ \left(  \frac{ 1}{2 } \omega_{ci} \right)^2}
 \left(1-\frac{1}{2} \frac{1}{L_n}  {\vec{u}}\cdot
 \vec{e}_x\right)
\label{approxfirstvelocitymoment}
\end{eqnarray}

\subsection{Boundary Conditions}

For stability reasons, specularly reflecting boundary conditions are chosen on the east and
west boundaries (in ``radial'' direction in terms of drift wave terminology),
whereas the north and south boundaries (in ``poloidal'' direction) are treated 
periodically. 
The rigid walls on east and west are set on the outermost lattice nodes, which
corresponds to an on-site reflection of the perpendicular components of $\bar f$. 
On east the distribution function is flipped according to $\bar f_3 \rightarrow
\bar f_1$,  $\bar f_6\rightarrow \bar f_5$, $\bar f_7 \rightarrow
\bar f_8$, and vice versa for the west boundary. 
Applying the boundary condition on $\bar f$ instead of $f$ introduces a small vorticity
source on the east and west boundaries, which can be circumvented by 
subtracting out the $F_i$ terms before updating the boundaries. 
However, this discrepancy had only minor impact on the present numerical results. 

\subsection{The CHM-LBM algorithm}

The LBM can be advanced in time by a four step algorithm, after an initialization of
the macroscopic fields has been carried out by either setting the initial
distribution and computing the macroscopic quantities, or by setting the
macroscopic initial fields. In the computations presented in section
\ref{sec:numexp} the latter initialization has been applied. 

The four-step time cycle includes:
\begin{enumerate}
 \item Update $\phi$ and ${\vec{u}}$ by
 eqs.~(\ref{approxzerothvelocitymoment}) and (\ref{approxfirstvelocitymoment})
 with the help of eq.~(\ref{phi_u_bar}) as a function of the previous
 $\bar f_{i}$ and/or the previous $\phi$ and ${\vec{u}}$; 
 \item Obtain $f_{i}^{eq}$ from eq.~(\ref{edfalg}) and $F_{i}$ from
   eq.~(\ref{forcingfunctionalg}) as
   functions of the updated $\phi$ and ${\vec{u}}$;
 \item Collide the particles using 
\begin{align}
\qquad \bar f_{i}^{*}({\vec{x}},t) =  \bar f_{i} ({\vec{x}},t)
+ \bar{A}\left( \bar f_{i} ({\vec{x}},t) - f_{i}^{eq} ({\vec{x}},t) \right) 
+\bar{\lambda} F_{i}  ({\vec{x}},t);
\end{align}
 \item Stream the particles to the adjacent nodes using
\begin{equation}
 \bar f_{i}({\vec{x}}+{\vec\xi}_{i} \delta t ,t + \delta t) =  \bar f_{i}^{*} ({\vec{x}},t);
\end{equation}
\dots and return to step (1).
\end{enumerate}

\section{Conventional finite difference scheme for the CHM \label{sec:chmmodel} }

The proposed LB model is cross-verified with a conventional finite difference scheme
which directly solves the CHM fluid eq.~(\ref{chm}), including an artificial hyperviscocity.

The employed Arakawa-Karniadakis scheme has first been applied to drift wave turbulence
computations by Naulin and Nielsen \cite{naulin03}, and uses the 3rd-order
accurate energy and enstrophy conserving Arakawa spatial discretization
\cite{arakawa66} for the Poisson bracket nonlinearity in combination with 
3rd-order ``stiffly stable'' time-stepping \cite{karniadakis91}.
The Karniadakis time-stepping scheme is here however reduced down to 2nd-order
to achieve the same temporal accuracy as in the present CHM-LBM algorithm. 
The electrostatic potential $\delta \phi$ is obtained after time stepping by solution of
the generalized Poisson Problem $(1-\nabla^2)\delta \phi = S$ with
a half-wave Fourier transform method. 
The boundary conditions are periodic in y direction and Dirichlet in x direction. 
In summary the global accuracy of the reduced Arakawa-Karniadakis method equals
the CHM-LB method and is of second order. 

\section{Numerical tests \label{sec:numexp}}

For all following computations the drift parameter is set to $\kappa_n = 0.05$, 
the normalized box size is fixed to ${L}_B = 64$, and the 
LB viscosity to ${\nu} =0.0002$. 

A stable LBM setup, which keeps the  $\mathcal{O}(\delta t)$
compressibility error as well as  the  $\mathcal{O}(\delta {x}^2)$
lattice error at a minimum level restricts their ratio $(\delta t/ \delta {x}^2) $ to a constant . 
A ratio of $(\delta t/ \delta {x}^2) \approx \mathcal{O}(1)$ performed with best stability especially in the decaying turbulence simulations
Hence the lattice resolution which fullfills this condition for the chosen parameters is  $N_x = 2048$ . 
The initial $E \times B$ drift velocity field is calculated with the help of
the lowest order momentum balance equation $\vec{e}_z \times {\vec{u} }=
(\kappa_n)^{-1} {\vec\nabla} \phi$. 
This guarantees that the potential field and the velocity field are consistent,
and hence initial pressure waves are supressed.
The finite difference scheme (FD) parameters differ from the LBM setup only by the
number of grid points $N_x = 512$ and the hyperviscous term of order 8 with 
viscosity parameter ${\nu}_8=10^{-9}$.  

\subsection{Monopole propagation}

The first electrostatic drift wave test case follows the evolution of Gaussian monopoles.
Within the framework of the CHM model, monopoles for various initial amplitudes $A$,
correponding to a rotation number $R_E = A/{r}_0$, exhibit nearly coherent
vortex propagation into the diamagnetic (here: $y$) direction for a large $R_E
\gg 1$, or contrarily, dispersive spreading for small $R_E \ll 1$  \cite{horton94,naulin02}. 

\begin{figure*}[!ht] 
\centering
 \caption*{\(t=0\)\hspace{37 mm} \(t=13\) \hspace{37 mm} \(t=25\)}
  \includegraphics[trim = 130px 350px 120px 130px, clip , scale=0.37]{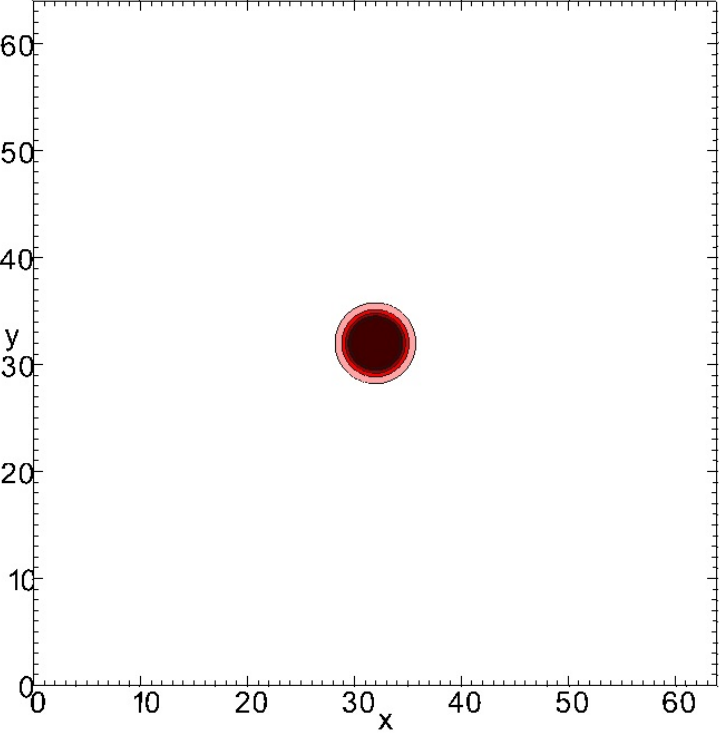}  
   \includegraphics[trim = 120px 350px 120px 130px, clip, scale=0.37]{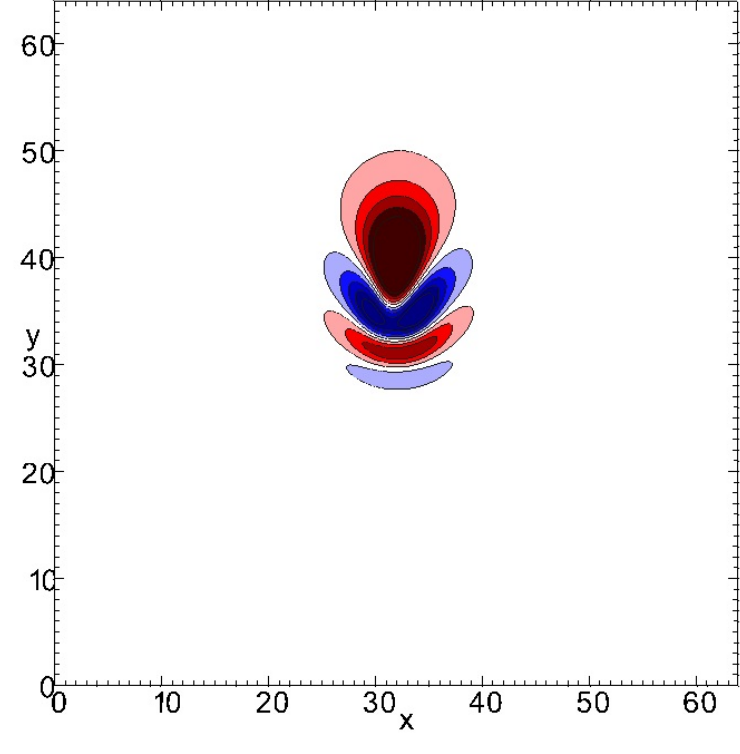}
   \includegraphics[trim = 120px 350px 120px 130px, clip, scale=0.37]{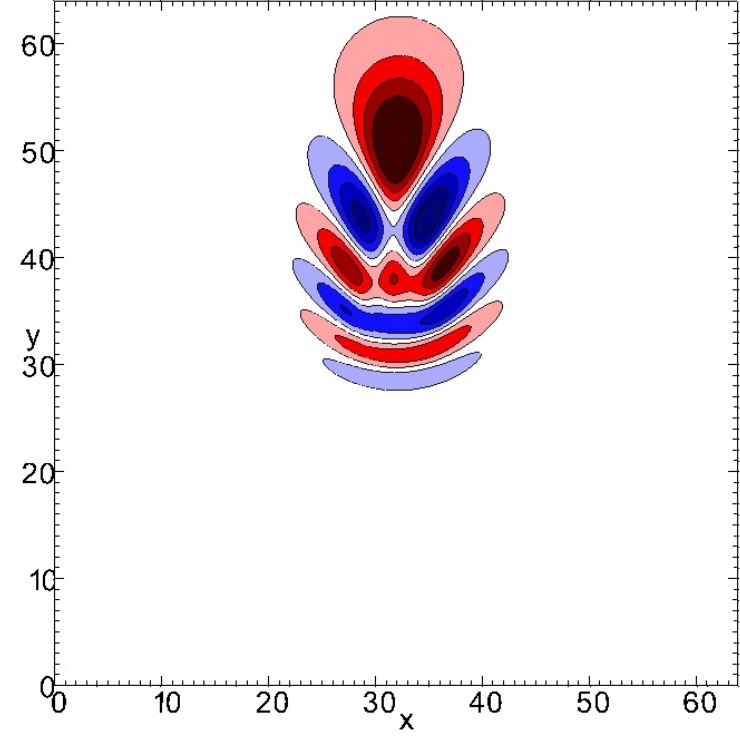}
   \includegraphics[trim = 250px 385px 270px 160px,clip,scale=0.45]{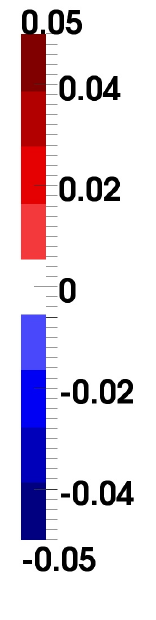}

  \includegraphics[trim = 85px 325px 107px 104px, clip, scale=0.32]{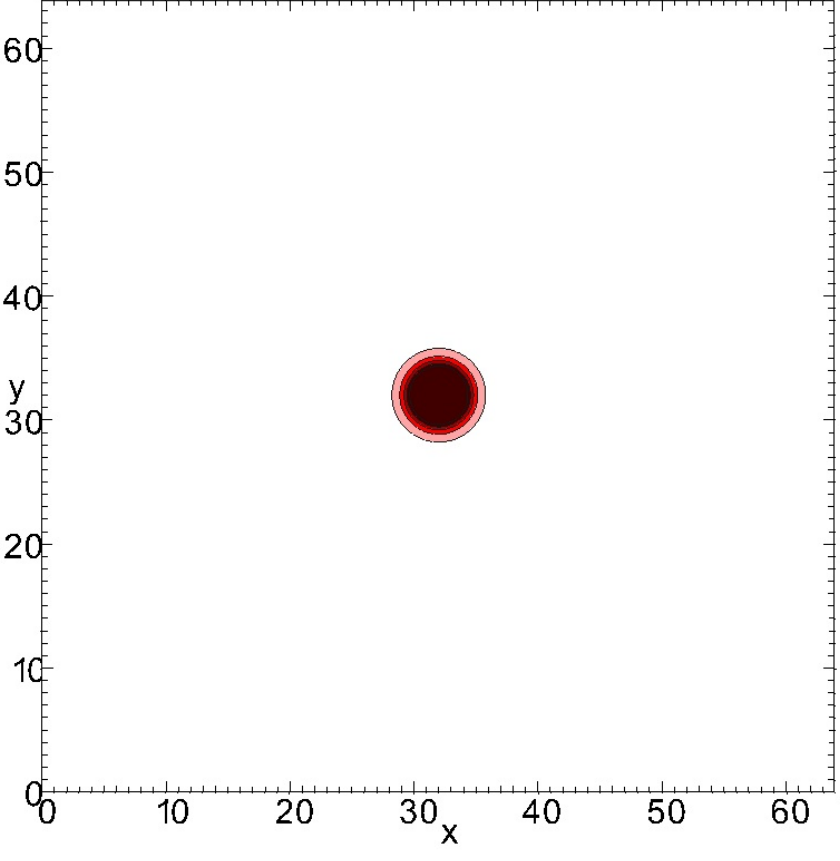}
  \includegraphics[trim = 77px 325px 107px 104px, clip, scale=0.32]{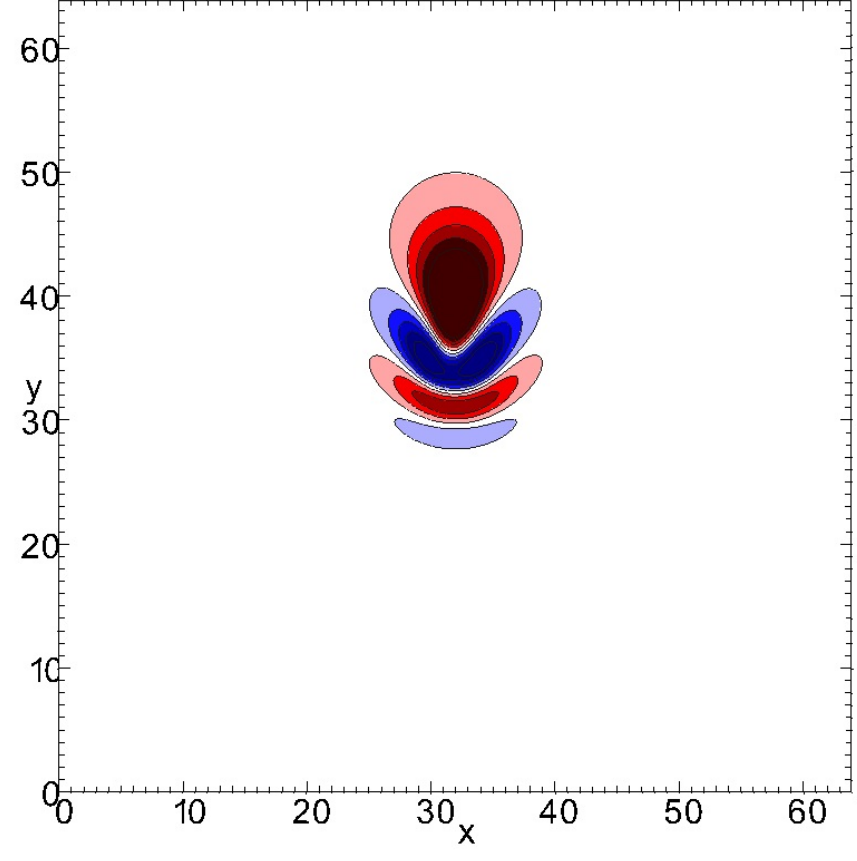}
  \includegraphics[trim = 75px 325px 107px 104px, clip, scale=0.32]{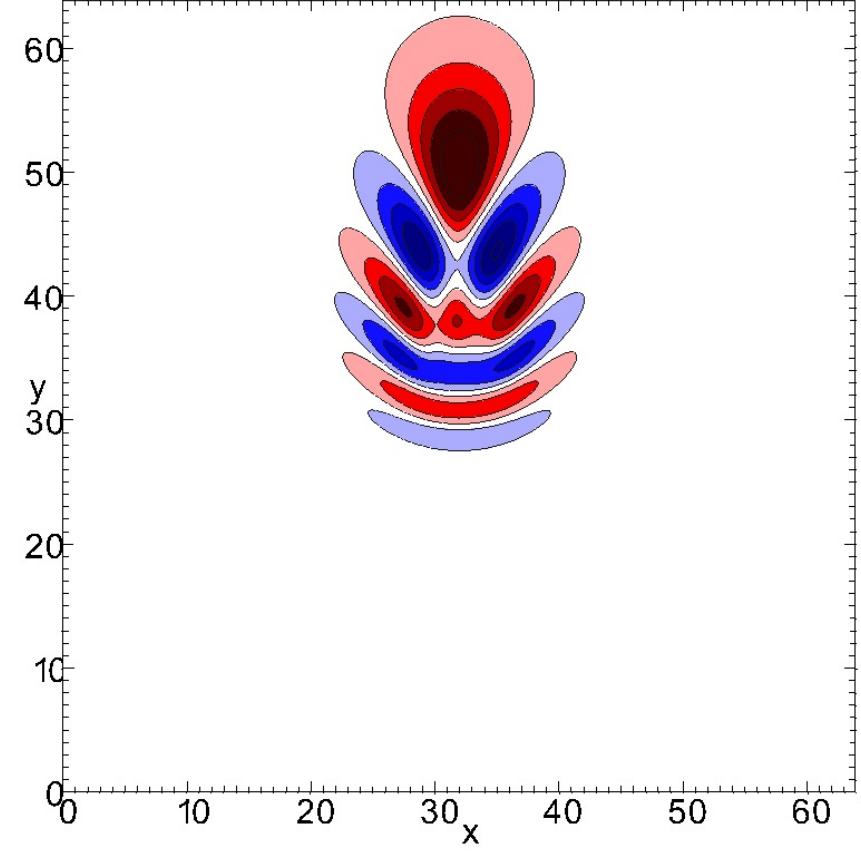}
  \includegraphics[trim = 250px 385px 270px 160px,clip,scale=0.45]{colorlegend_0_05.pdf}
\caption{\label{fig:mono01} Evolution of an initial Gaussian monopole for 
$R_E  = 0.1$: the equally spaced isolines represent the potential field $\delta \phi$
  of the LBM (first row) and the FD scheme (second row) at various times.} 
  
%................
\end{figure*}

\begin{figure*}[!ht] 
\centering
 \caption*{\(t=0\)\hspace{37 mm} \(t=13\) \hspace{37 mm} \(t=25\)}
     \includegraphics[trim = 70px 290px 120px 116px, clip, scale=0.32]{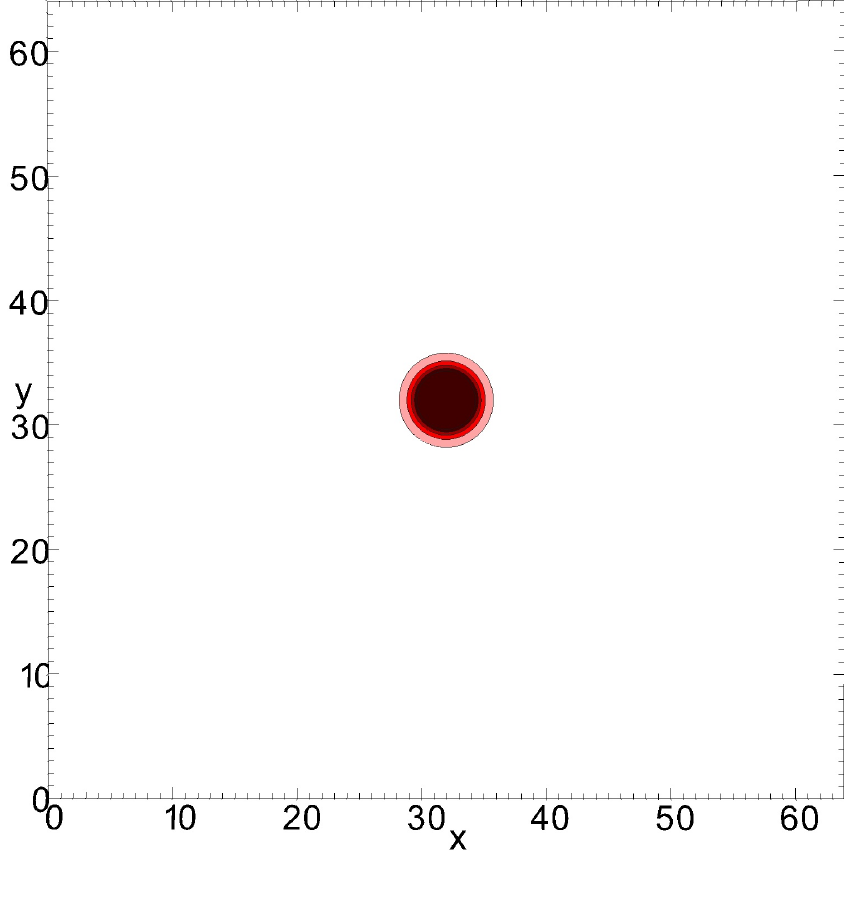}
     \includegraphics[trim = 75px 290px 120px 116px, clip, scale=0.32]{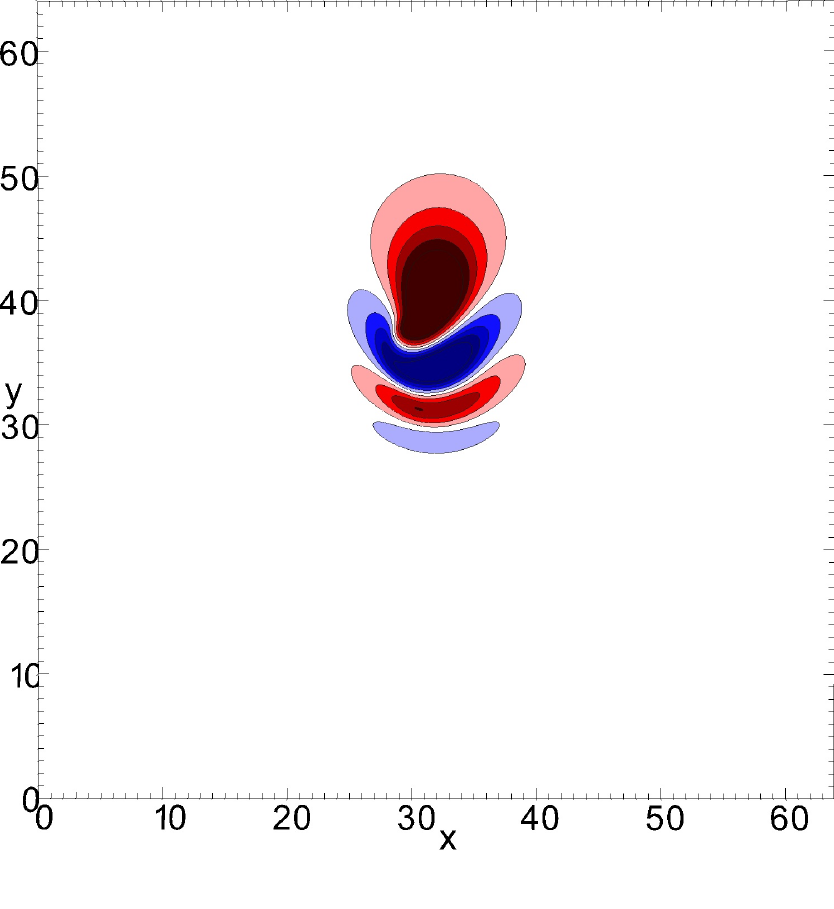} 
     \includegraphics[trim = 75px 290px 120px 116px, clip, scale=0.32]{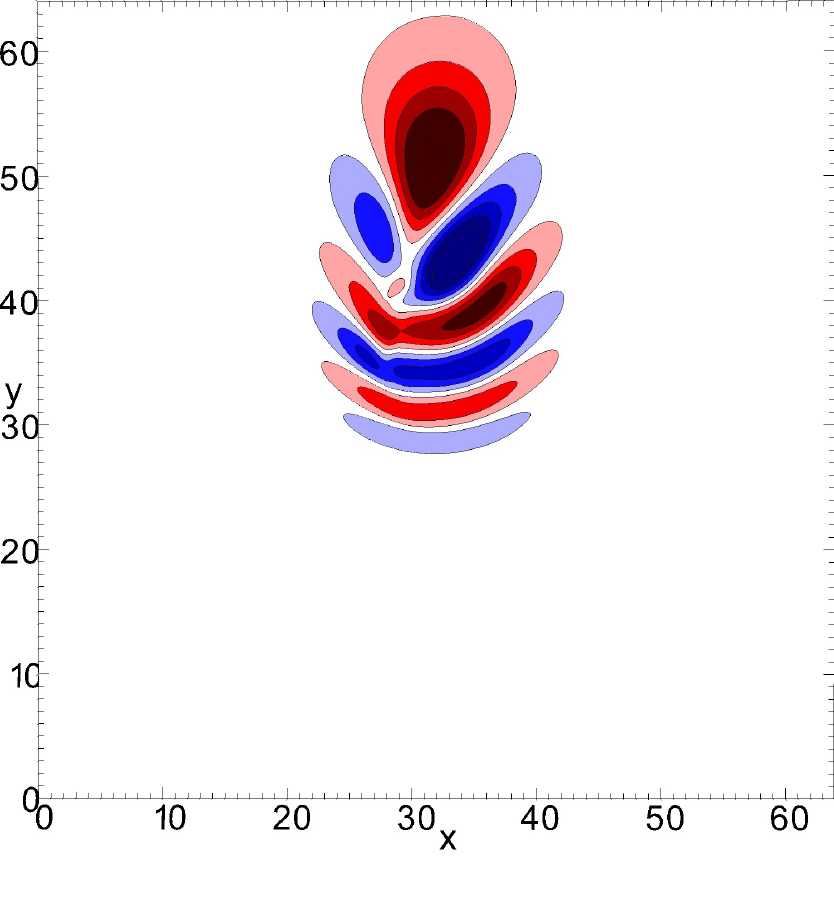}
     \includegraphics[trim = 250px 370px 290px 160px,clip,scale=0.45]{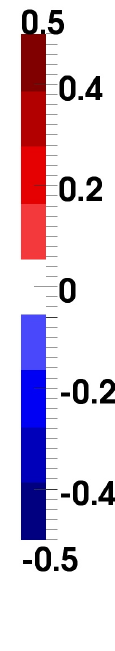}
       
    \includegraphics[trim = 70px 305px 104px 104px, clip, scale=0.31]{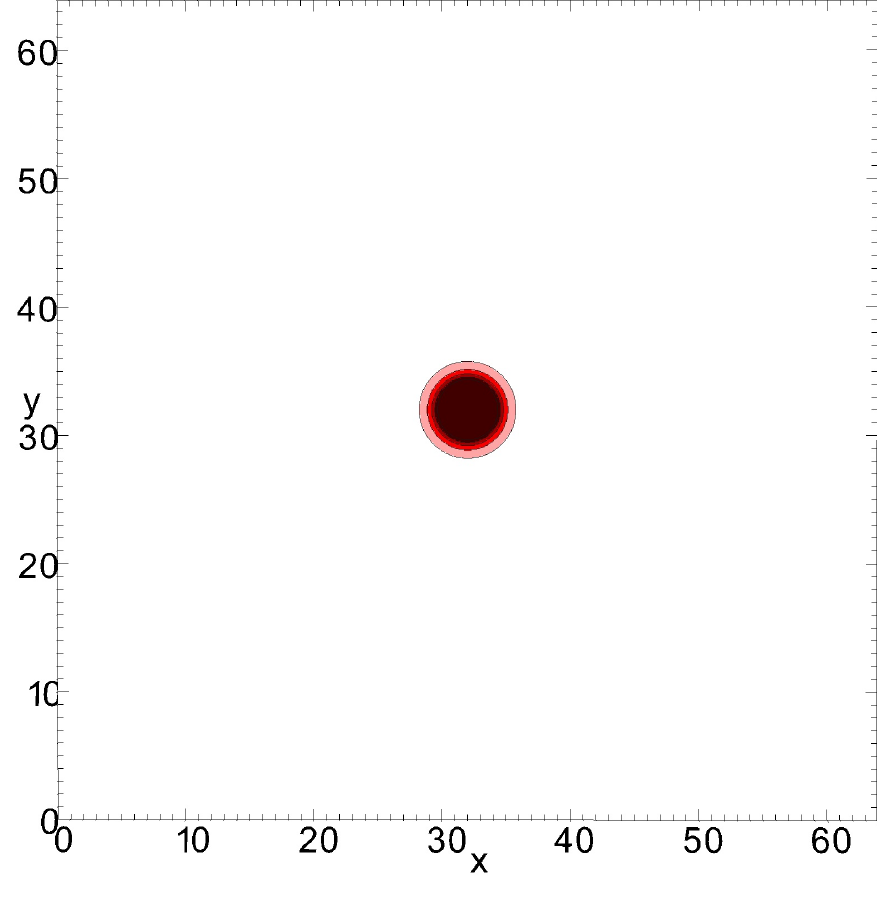}
    \includegraphics[trim = 75px 305px 104px 104px, clip, scale=0.31]{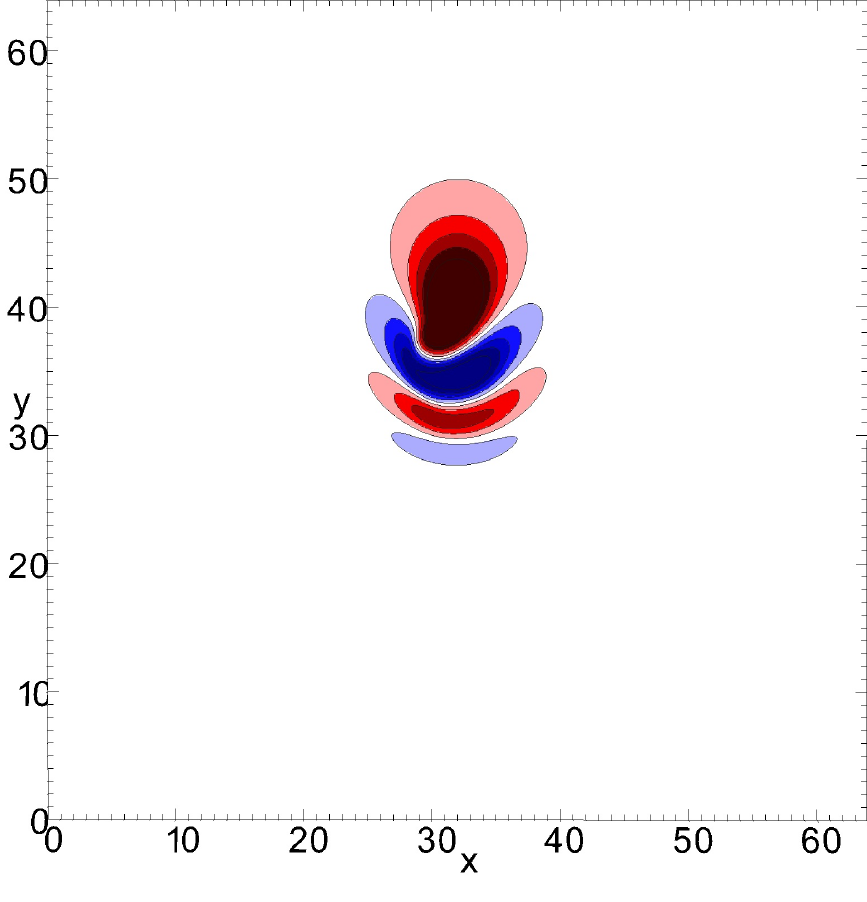}
    \includegraphics[trim = 75px 305px 104px 104px, clip, scale=0.31]{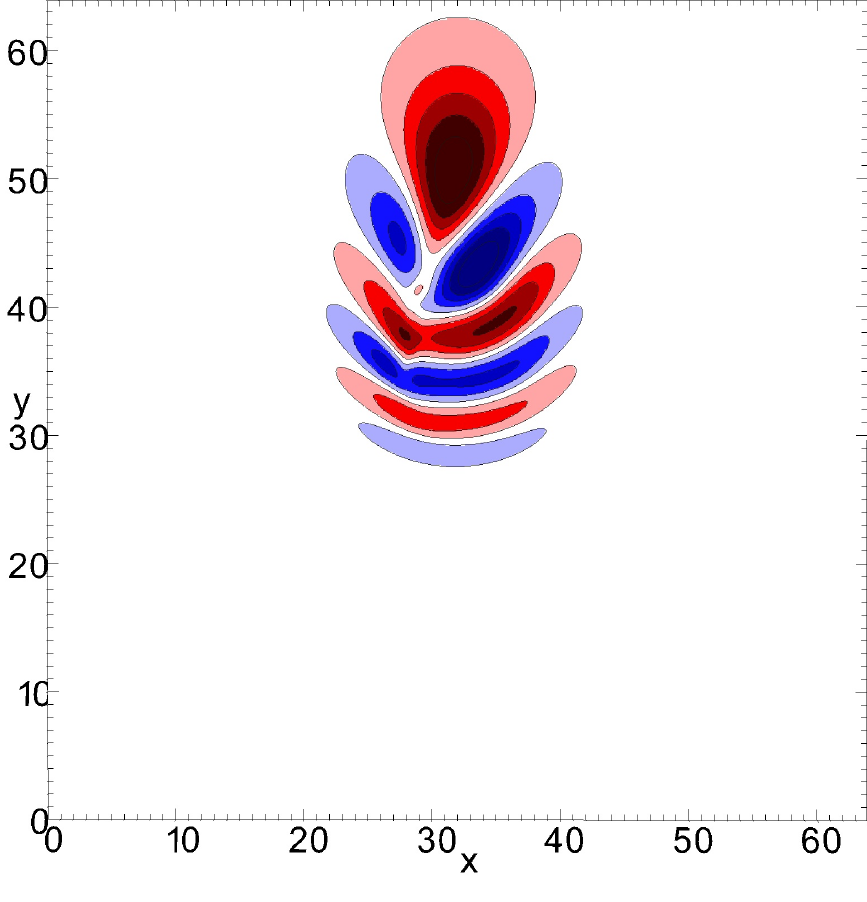}
    \includegraphics[trim = 250px 385px 290px 160px,clip,scale=0.45]{colorlegend_0_5.pdf}
 \caption{\label{fig:mono1}Evolution of an initial Gaussian monpole for 
$R_E  = 1$: Self focusing is concurring with monopole spreading and results in
  a mixed form of the linear and nonlinear regime.}  
  
%................
\end{figure*}

\begin{figure*}[!ht]
 \centering
 \caption*{\(t=0\)\hspace{37 mm} \(t=13\) \hspace{37 mm} \(t=25\)}
	\includegraphics[trim = 40px 260px 85px 78px, clip, scale=0.28]{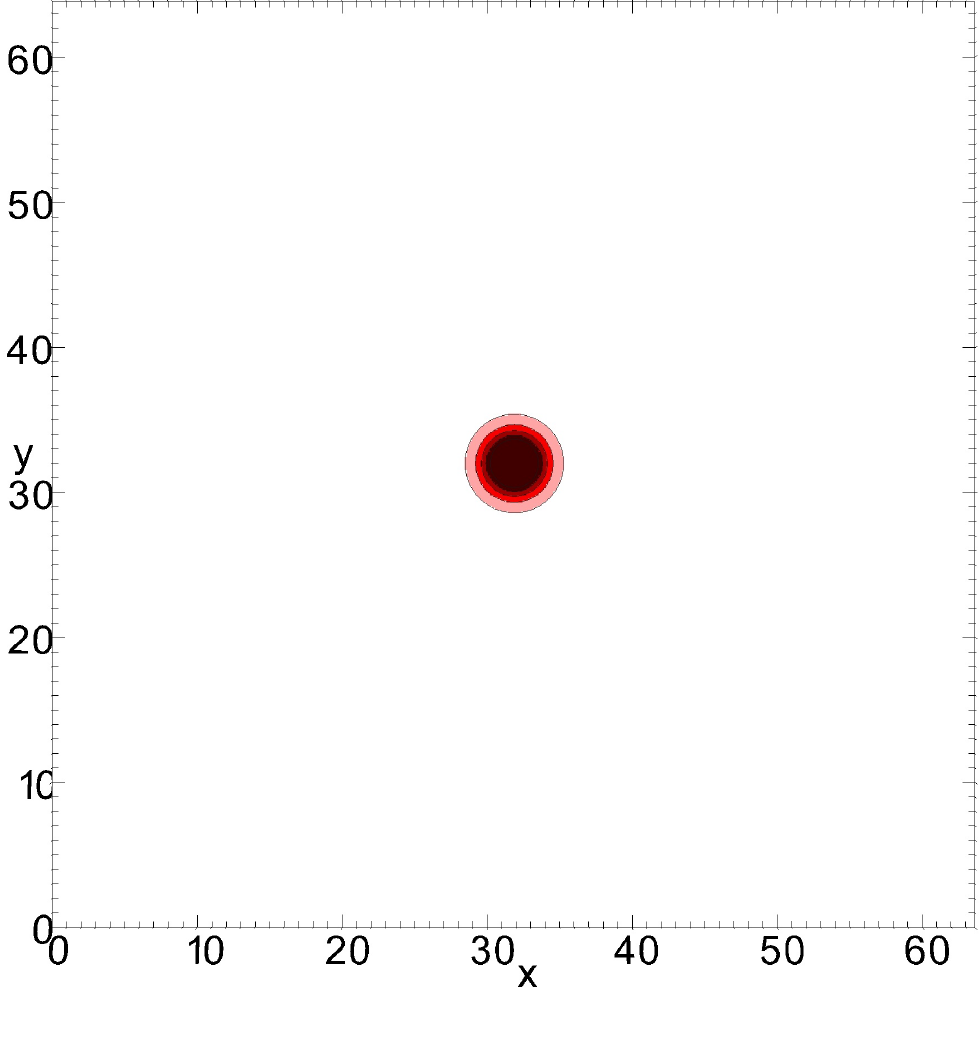}
	\includegraphics[trim = 40px 260px 85px 78px, clip, scale=0.28]{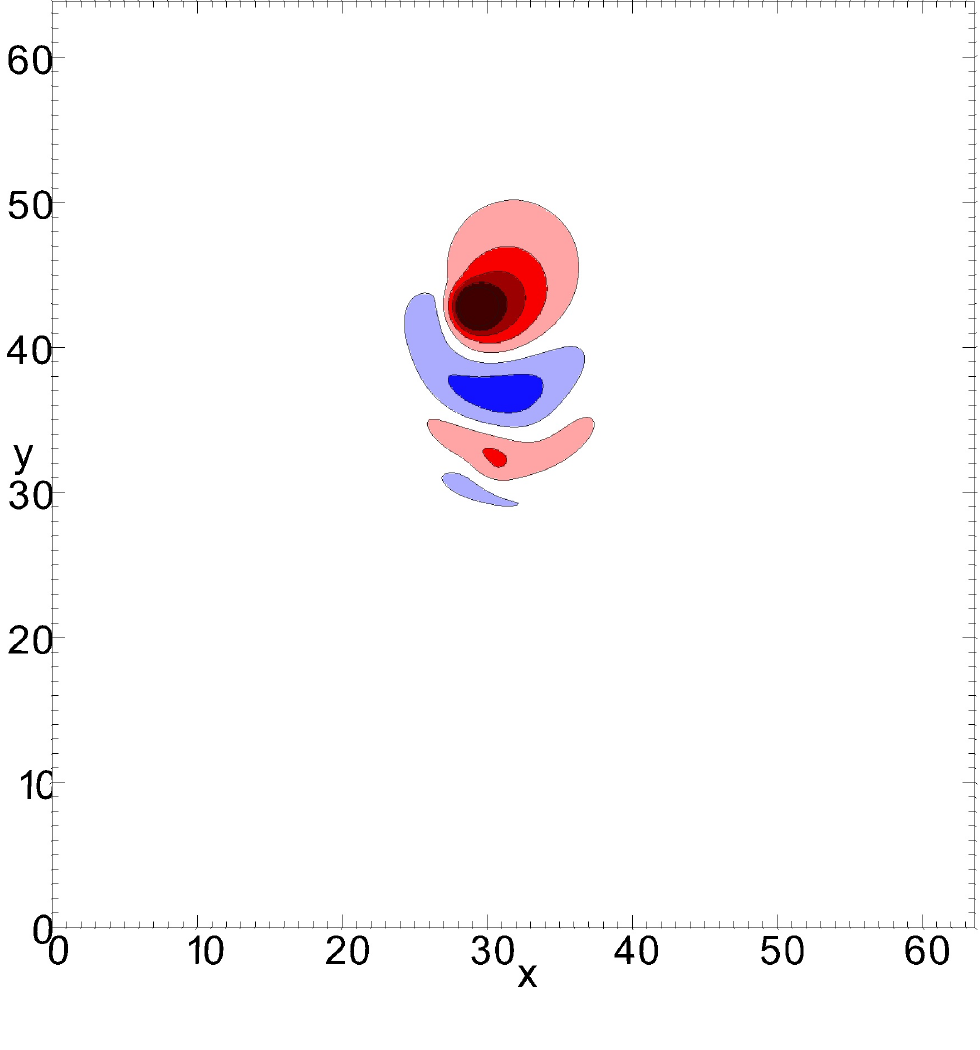} 
	\includegraphics[trim = 40px 260px 85px 78px, clip, scale=0.28]{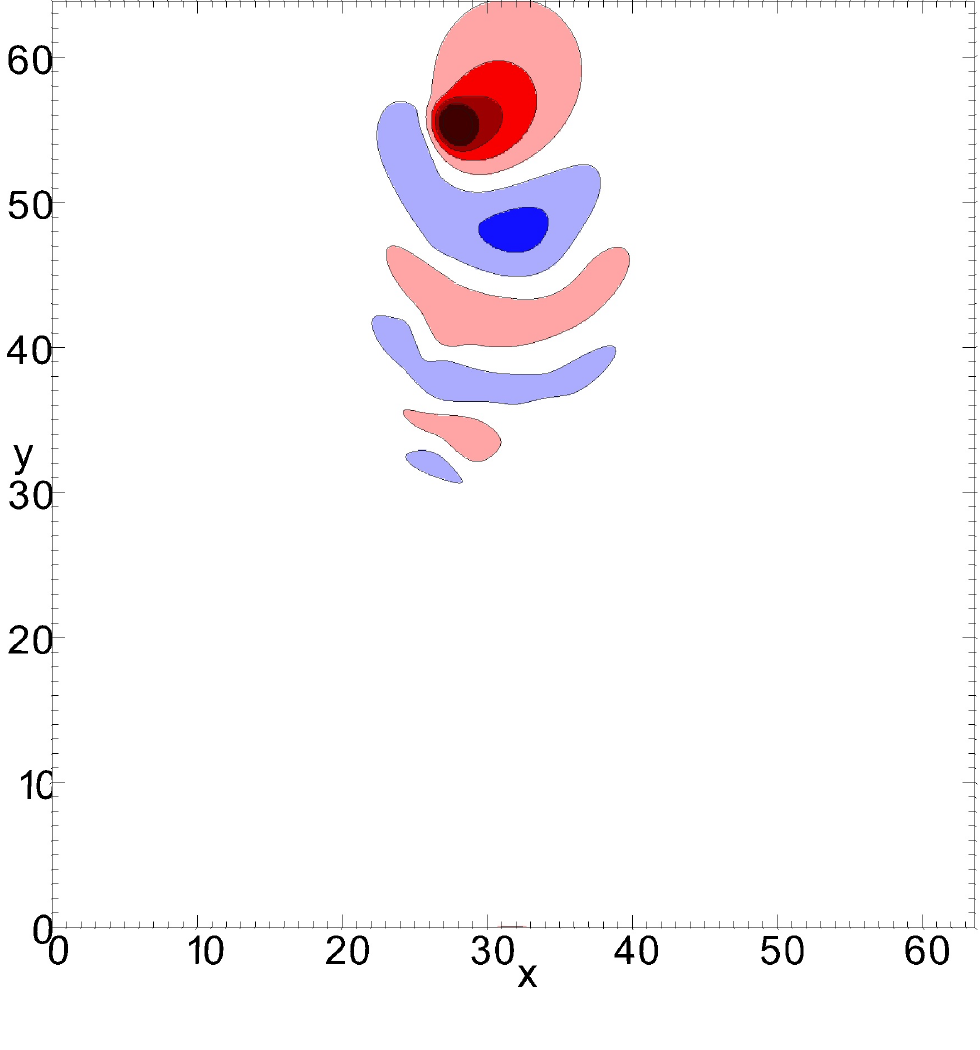}
	\includegraphics[trim = 250px 370px 290px 150px,clip,scale=0.45]{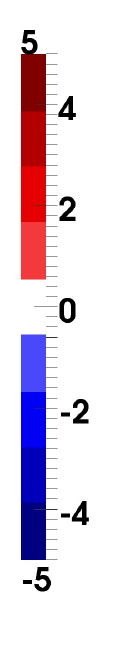}
	
	\includegraphics[trim = 25px 250px 65px 58px, clip, scale=0.26]{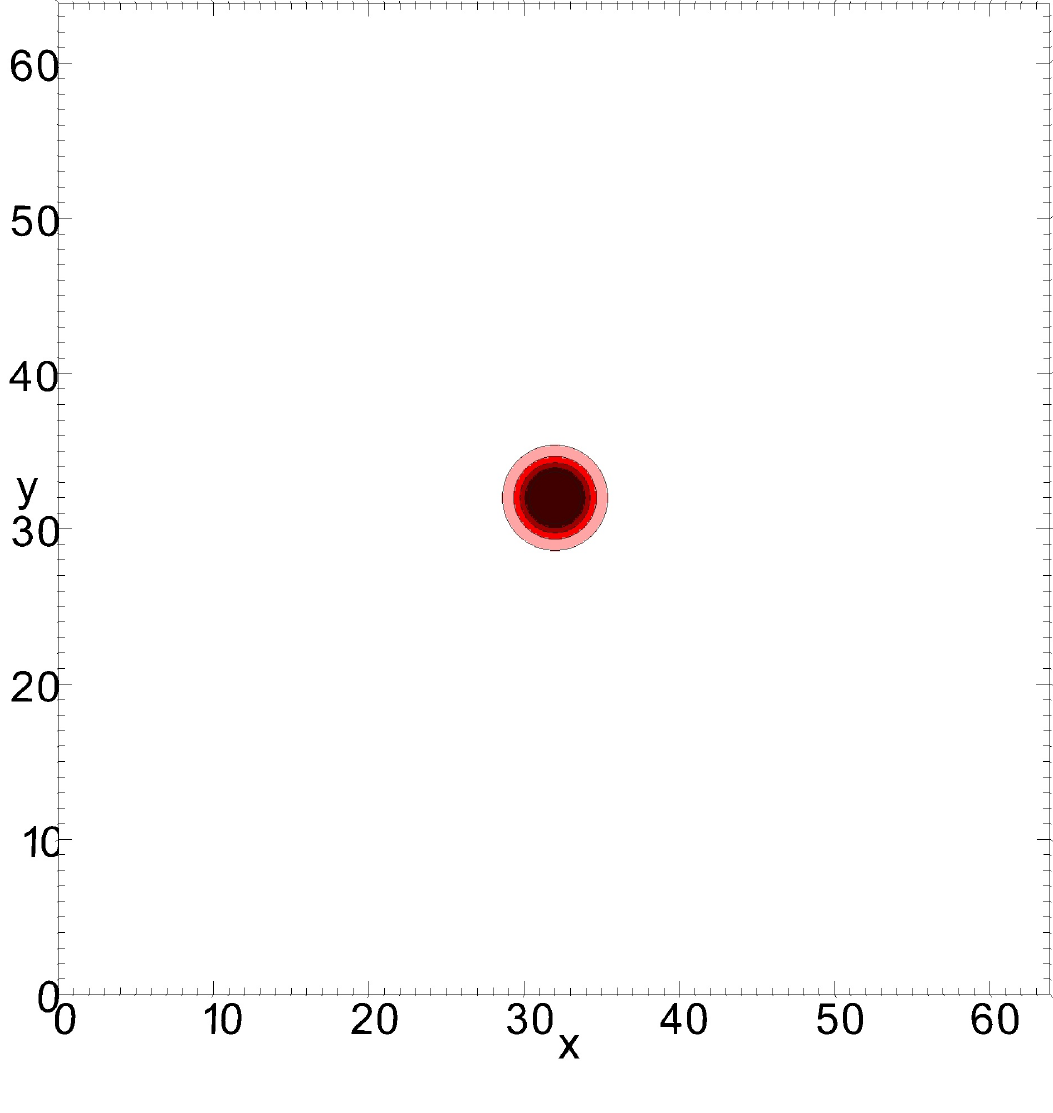}
	\includegraphics[trim = 25px 250px 65px 58px, clip, scale=0.26]{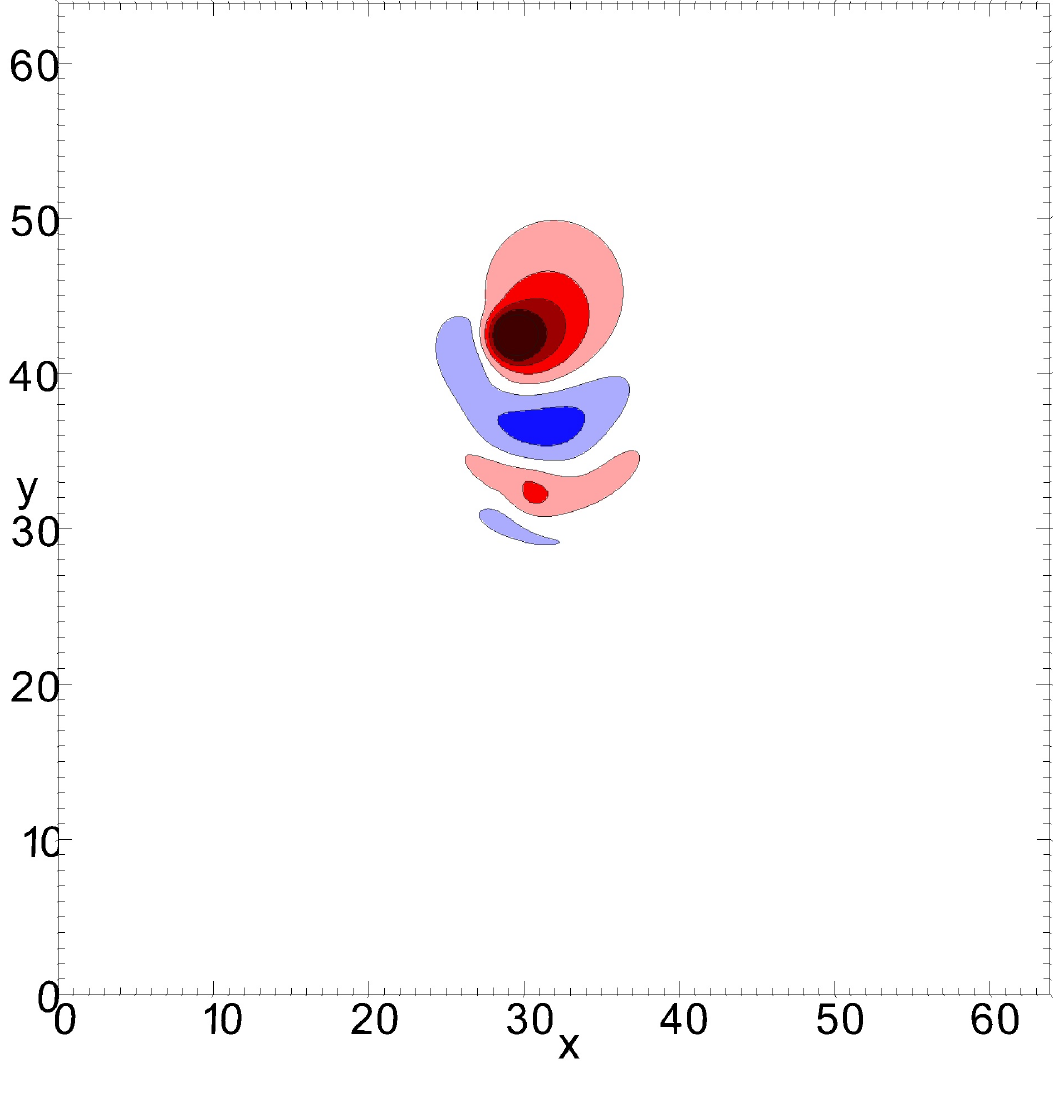}
	\includegraphics[trim = 25px 250px 65px 58px, clip, scale=0.26]{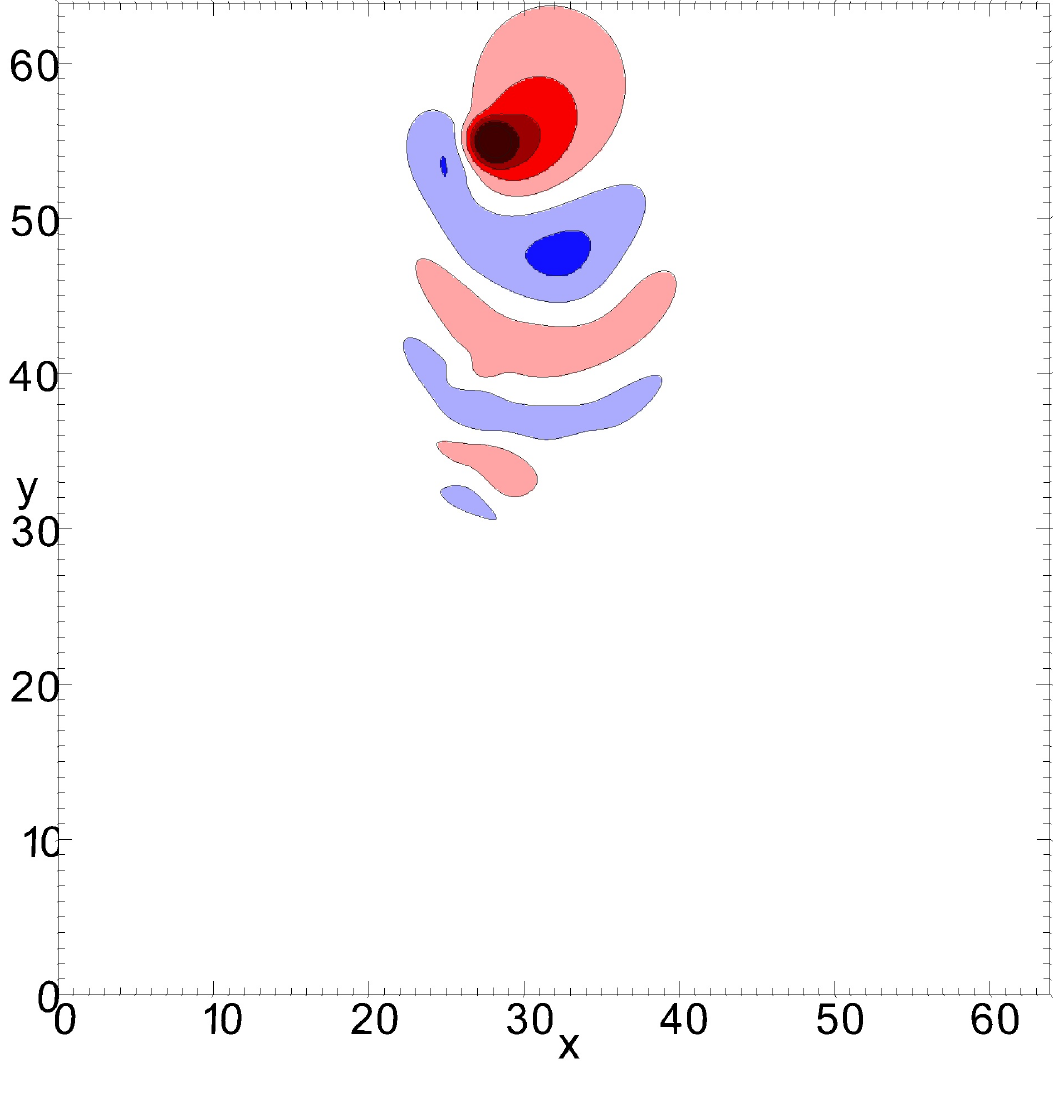}
       \includegraphics[trim = 250px 380px 290px 150px,clip,scale=0.45]{colorlegend_5.pdf}
\caption{\label{fig:mono5}Evolution of an initial Gaussian monpole for 
$R_E  = 5$: The dominant process is self-focusing, which allows the
  monopole to propagate nearly dispersionless with the drift velocity. (first row:
  LBM; second row: FDM)} 
\end{figure*}

In the following test cases, the propagation of initial monopoles with $R_E =
0.1$ (Fig.~\ref{fig:mono01}), $R_E = 1$ (Fig.~\ref{fig:mono1}) and $R_E = 5$ 
(Fig.~\ref{fig:mono5}) is compared between the LB (first row) and FD (second row)
schemes at various times of the computation.
The LB algorithm closely resembles the monopole dynamics of the CHM equation posed by
the FD scheme, except for a small deviation of the vortex amplitudes
at later times (compare Fig.~\ref{fig:mono01} at $t=25$). 

%................

\subsection{Dipole drift modons}

\begin{figure*}[!ht]
 \centering
 \caption*{\(t=0\)\hspace{37 mm} \(t=18\) \hspace{37 mm} \(t=36\)}
	\includegraphics[trim = 30px 245px 72px 63px, clip, scale=0.26]{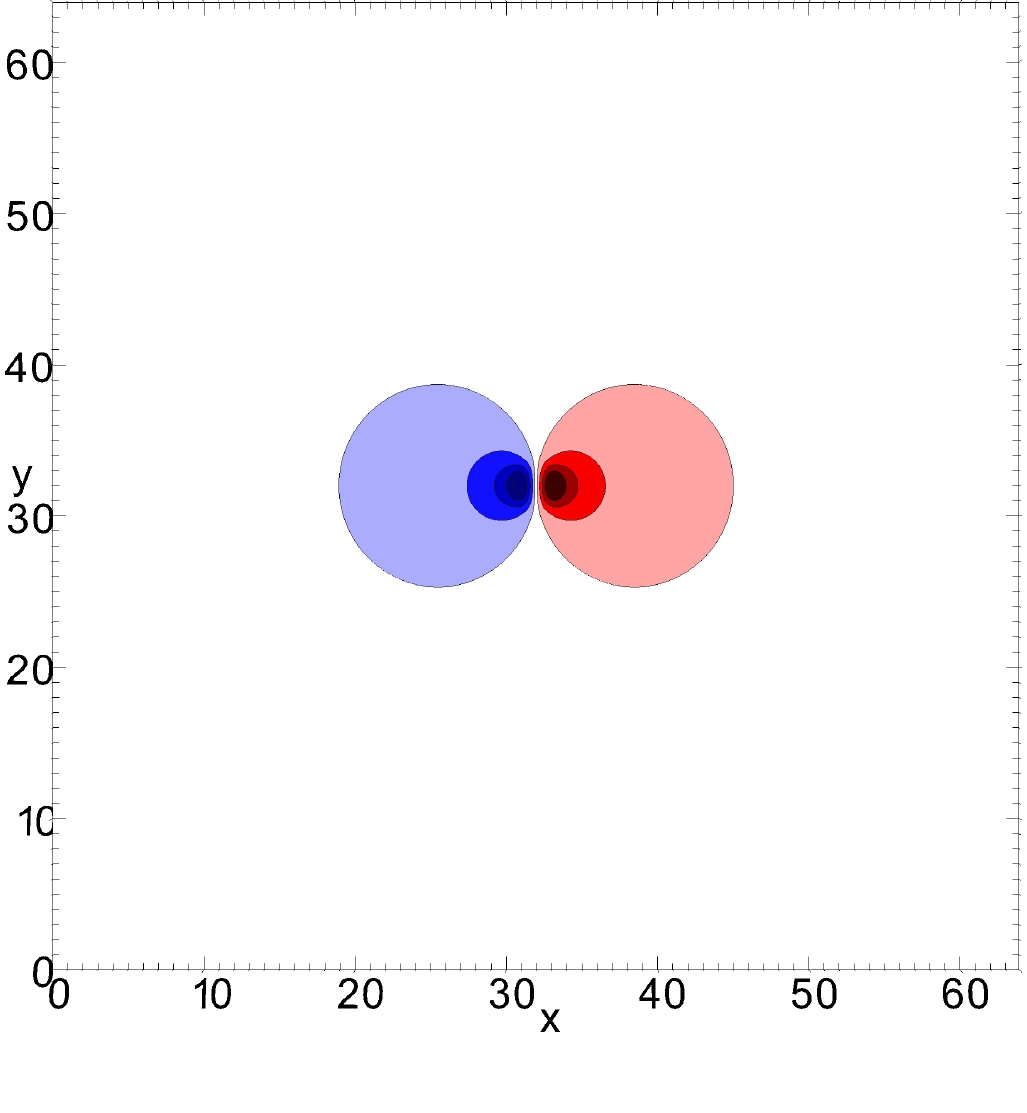}
	\includegraphics[trim = 30px 245px 72px 63px, clip, scale=0.26]{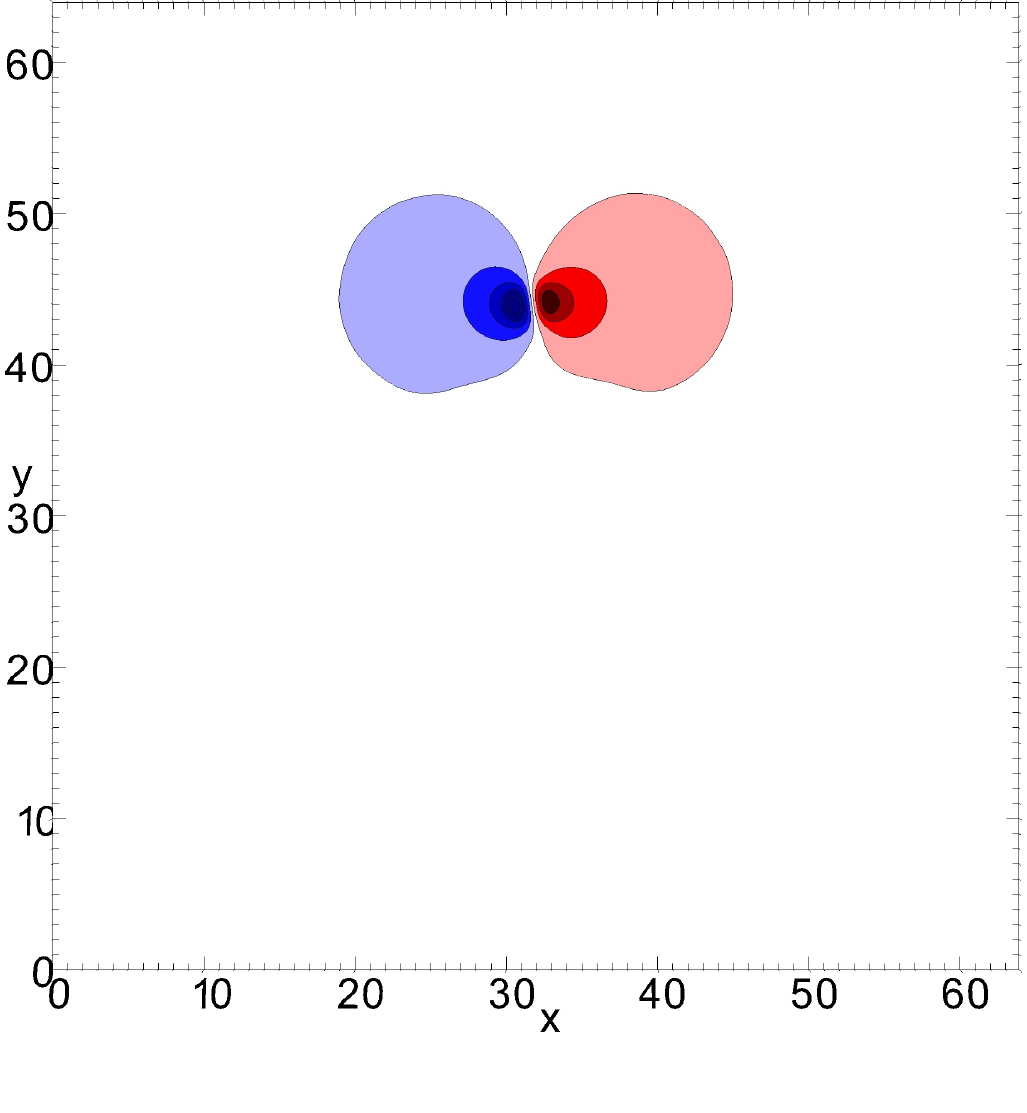} 
	\includegraphics[trim = 30px 245px 72px 63px, clip, scale=0.26]{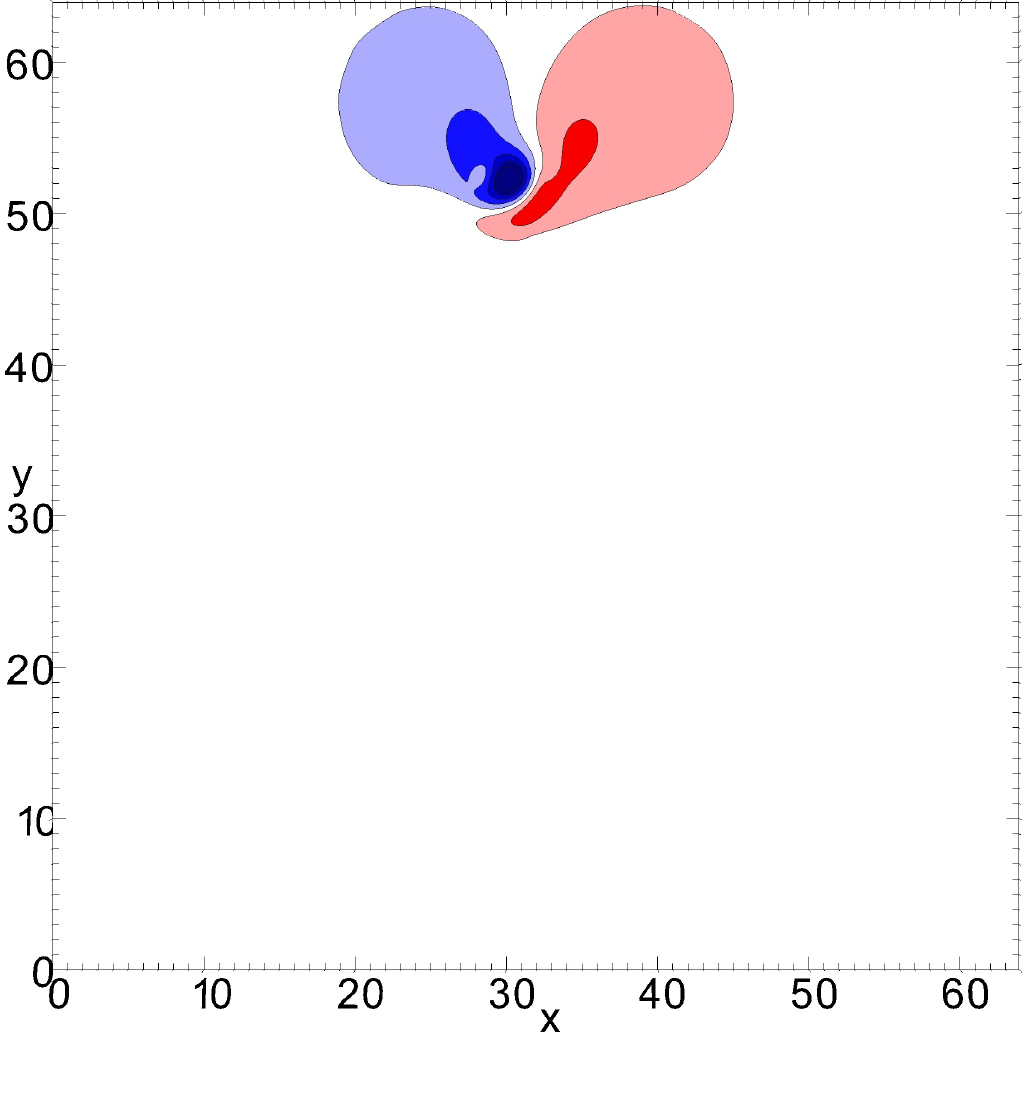}
	\includegraphics[trim = 250px 370px 290px 160px,clip,scale=0.45]{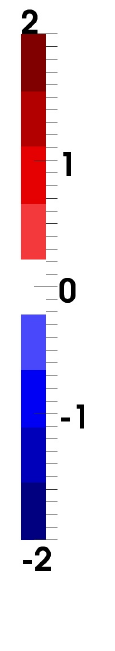}

	\includegraphics[trim = 50px 270px 88px 83px, clip, scale=0.28]{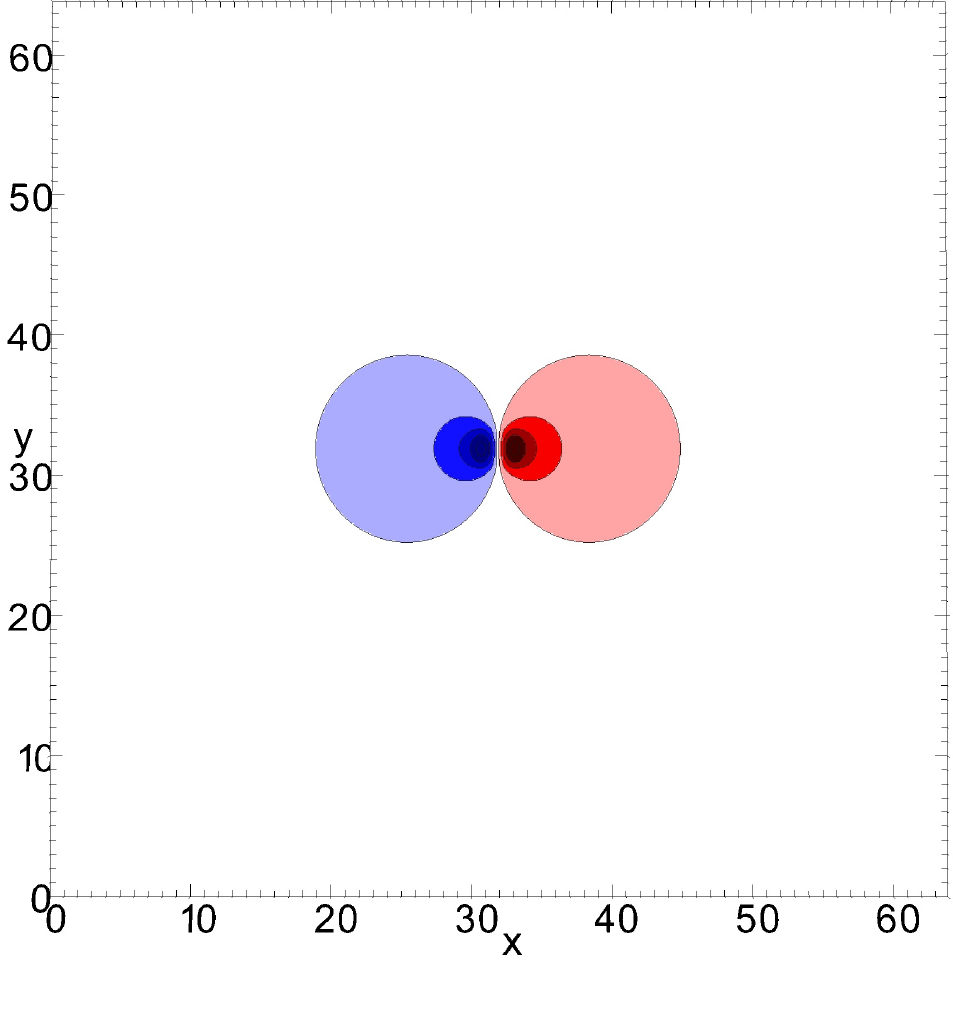}
	\includegraphics[trim = 50px 270px 88px 83px, clip, scale=0.28]{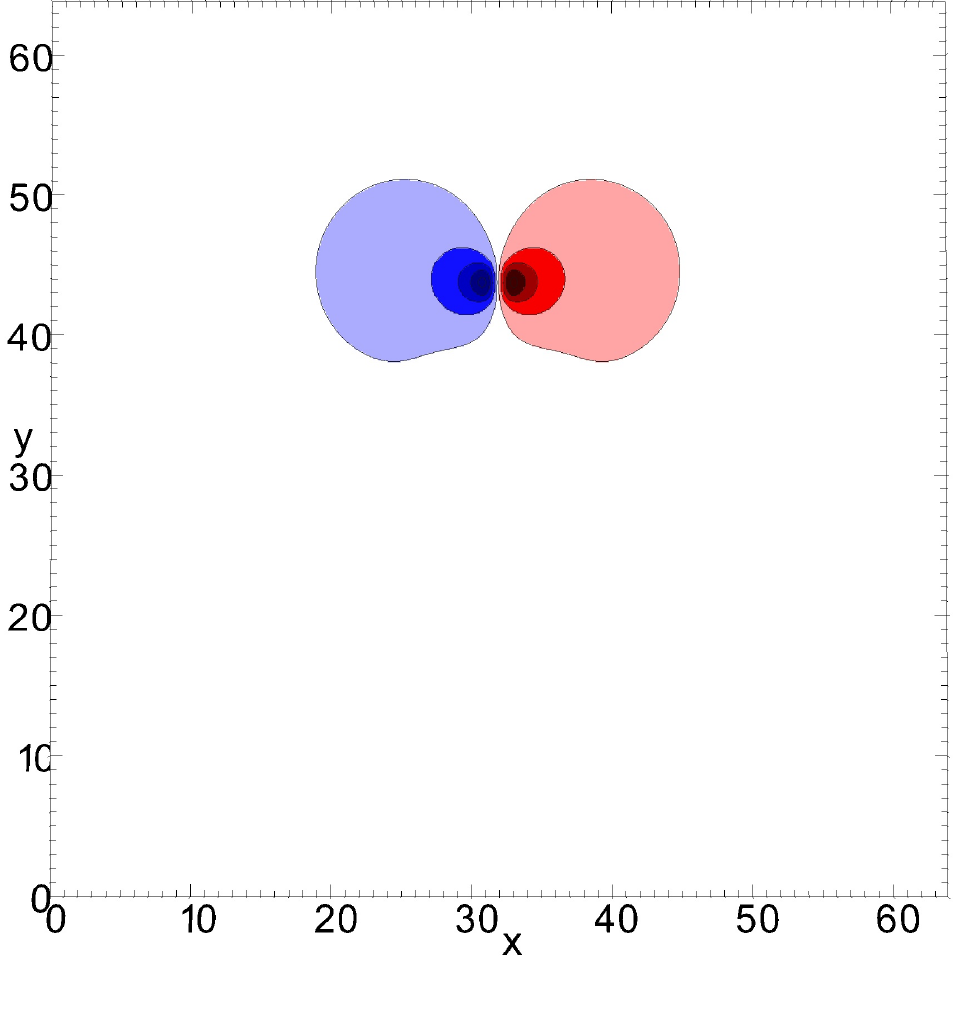}
	\includegraphics[trim = 50px 270px 88px 83px, clip, scale=0.28]{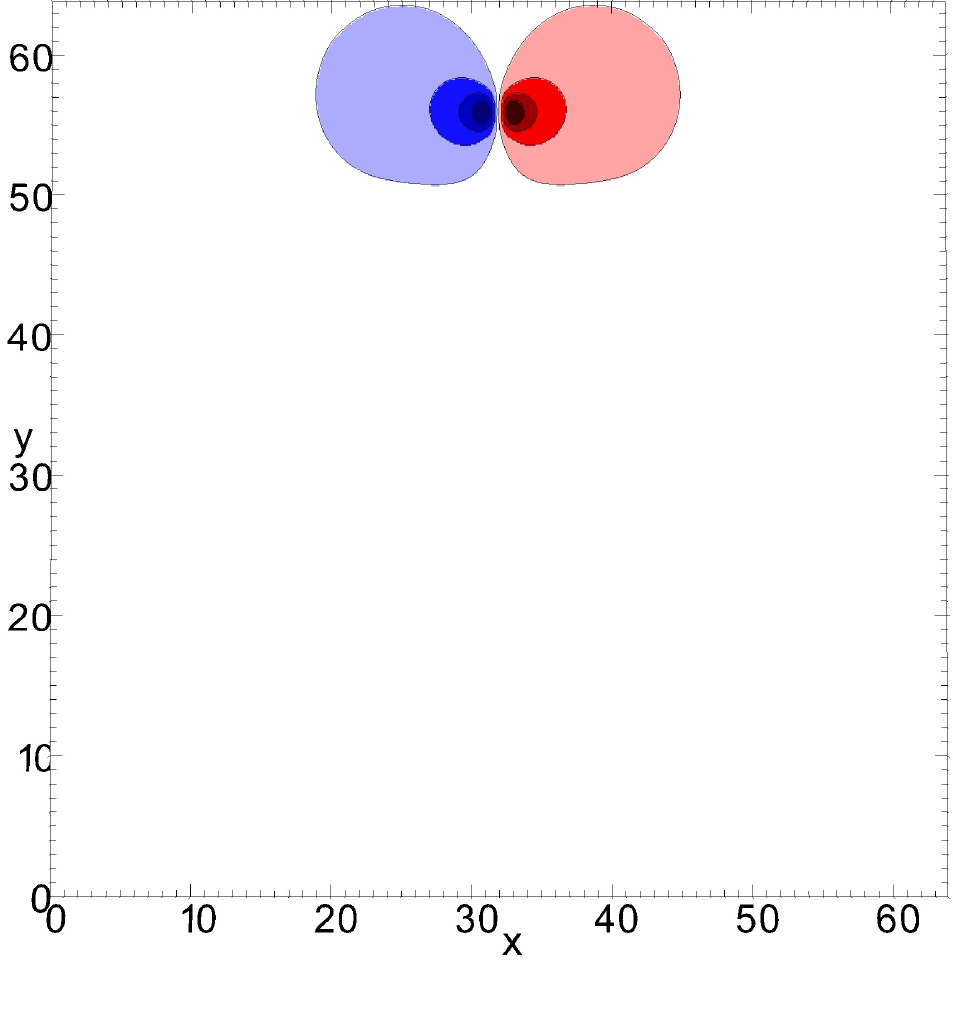}
	\includegraphics[trim = 250px 380px 290px 160px,clip,scale=0.45]{colorlegend_2.pdf}
\caption{\label{fig:dm_stable} The potential $\delta \phi$ of a stably
  propagating drift modon is shown for the LBM (first row) and for the FD
  scheme (second row). It is clearly visible that for the LBM case the shape of the
  modon is distorted by a small shear from around $t=36$ on.}
\end{figure*}

Solitary dipole drift vortex solutions or modons are, in contrast to Gaussian
monopoles, localized stationary solutions to the CHM equations. 
For Larichev-Reznik modons, the initial potential perturbation of radial extent
${R}$ is defined as \cite{fontan95,horton94}:

\begin{equation}
 \delta \phi({r},\vartheta)=  \phi_r ({r})  \cos{(\vartheta)} 
\end{equation}
with 
\begin{equation}
 \phi_r ({r}) =
 \begin{cases}
{u}_{*} {r} \left(1 +
  \frac{\beta^2}{\gamma^2} \right) - \frac{{u}_{*} {R} \beta^2
    J_1(\gamma {r})} {\gamma^2 J_1(\gamma {R})} &, {r}\le {R} \\
 \frac{{u}_{*} {R} K_1(\beta {r})}{K_1(\beta {R})}  &, {r}>{R}
 \end{cases}
\end{equation}
where $J_1$, $J_2$ are Bessel functions of the first kind and $K_1$, $K_2$ are
modified Bessel functions of the second kind. The parameter $\beta =
\sqrt{1- ({u}_d / {u}_{*})}$  contains the ratio of the drift
velocity ${u}_d$ to the dipole velocity ${u}_{*}$. 
The parameter $\gamma$ is determined by the transcendental equation
\begin{equation}
 \frac{K_2(\beta {R})}{\beta {R} K_1(\beta {R})} = - \frac{J_2(\gamma {R})}{\gamma {R} K_1(\gamma {R})}
\end{equation}
Its smallest number defines the ground state of the dipole, and higher order
solutions are excited states \cite{kono10}. 
The drift modon is stable if the ratio between the typical dipole velocity to
the drift velocity fullfills ${u}_d / {u}_{*} \ge 1$ whereas dispersive 
broadening appears for negative ratios. The present computations are restricted to the ground
state of stable travelling drift modon with ${u}_d / {u}_{*}  =2$. 
Fig.~\ref{fig:dm_stable} shows the expected propagation of the drift
modon into the $+y$ direction. For the LB model (first row) an increased shearing of the
drift modon is observed in comparison to the classical FD model (second row)
with progressing time. 
In the LB model the lifetime of the stable travelling drift modon is 
further enhanced by reducing the drift ratio \(\kappa_n\).

%................

\subsection{Decaying Turbulence}

\begin{figure*}[!ht]
 \centering
 \caption*{\(t=0\)\hspace{37 mm} \(t=27\) \hspace{37 mm} \(t=90\)}
    \includegraphics[trim = 30px 295px 82px 68px, clip, scale=0.27]{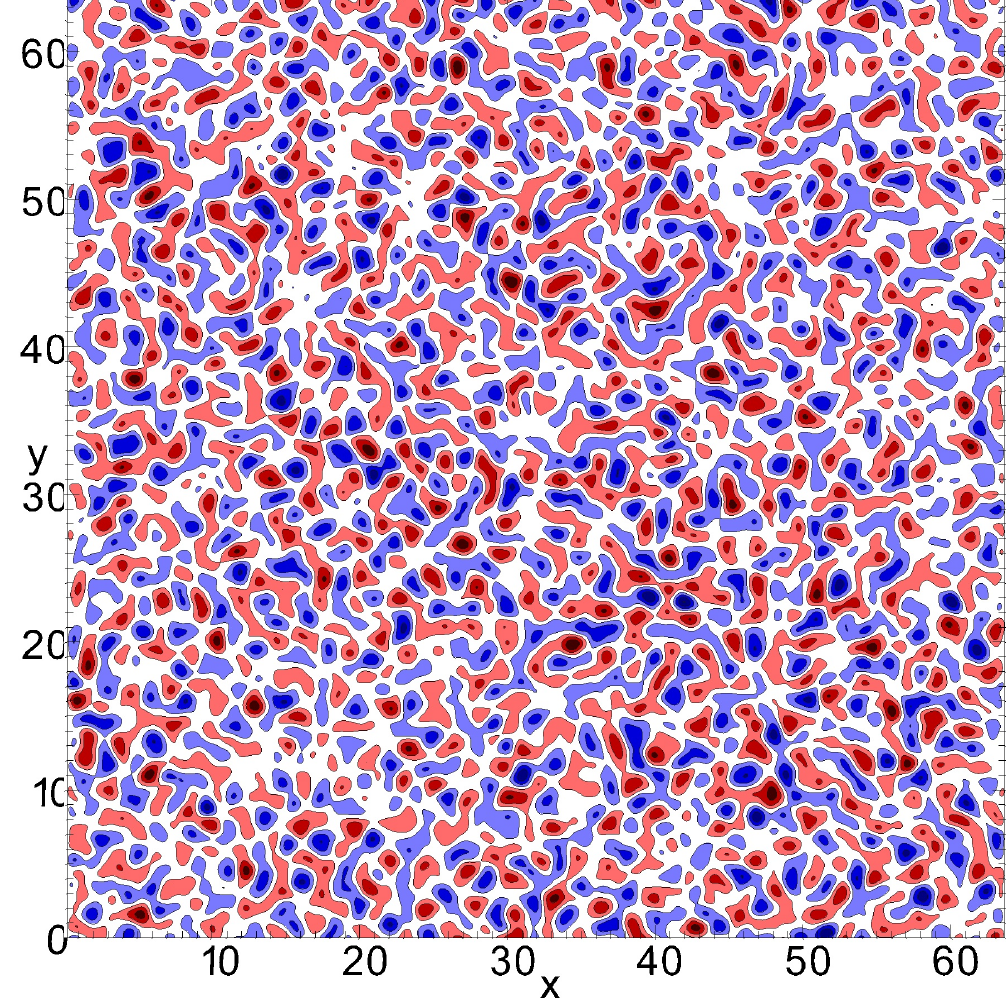}
    \includegraphics[trim =50px 295px 70px 68px, clip, scale=0.27]{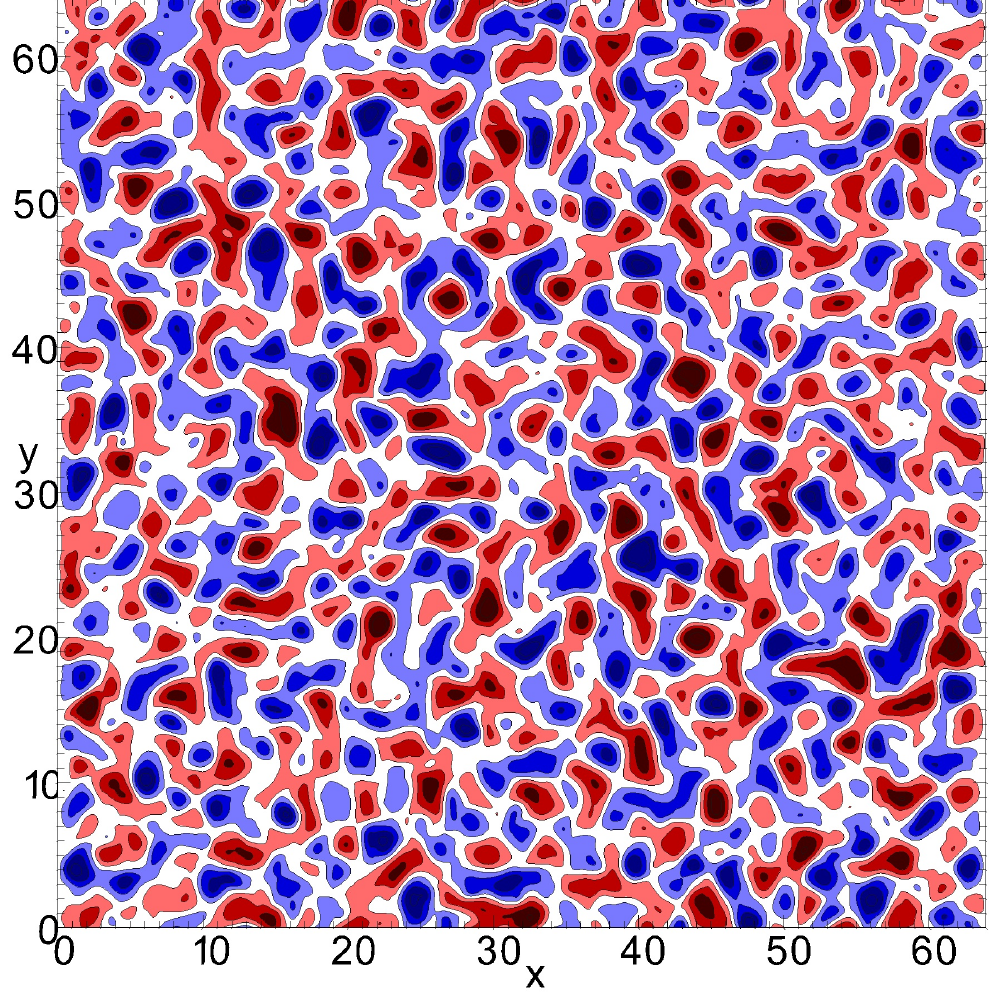}
    \includegraphics[trim = 35px 295px 78px 68px, clip, scale=0.27]{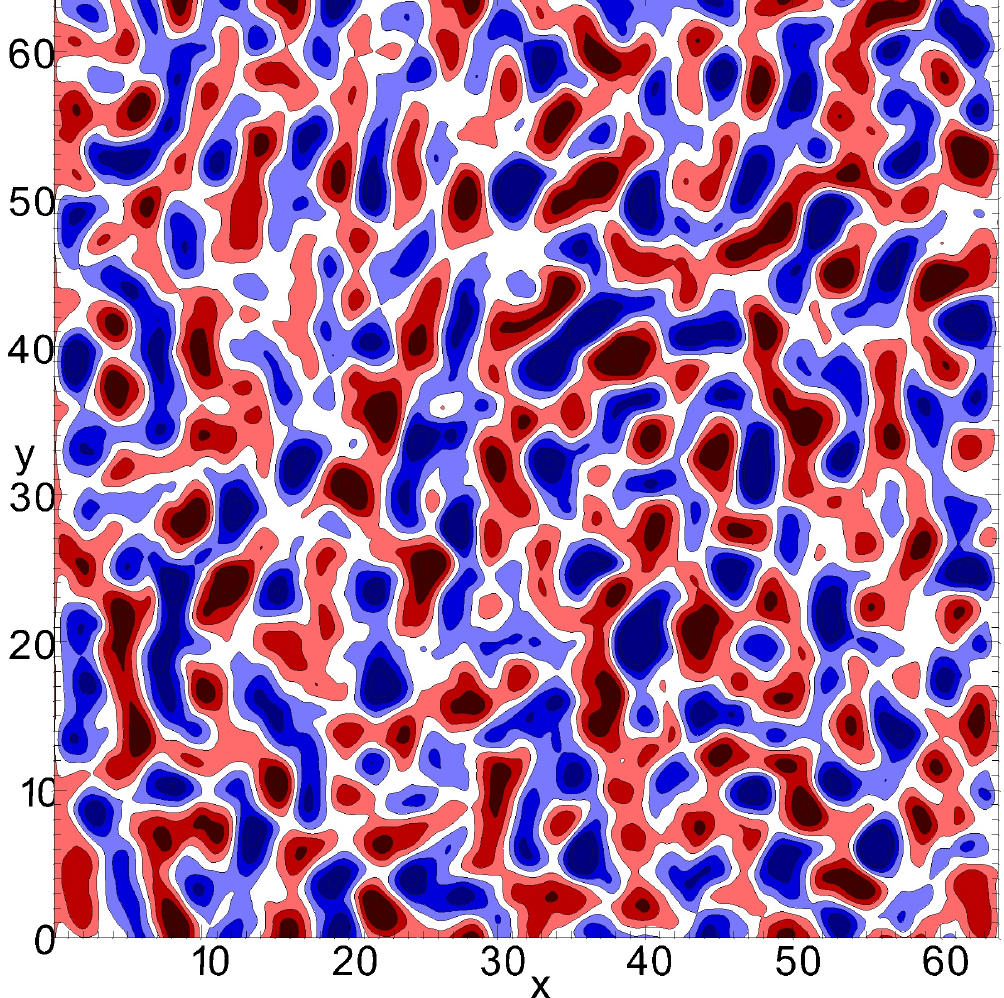}
     \includegraphics[trim = 250px 395px 290px 160px,clip,scale=0.45]{colorlegend_0_5.pdf}

    \includegraphics[trim = 30px 260px 72px 73px, clip, scale=0.27]{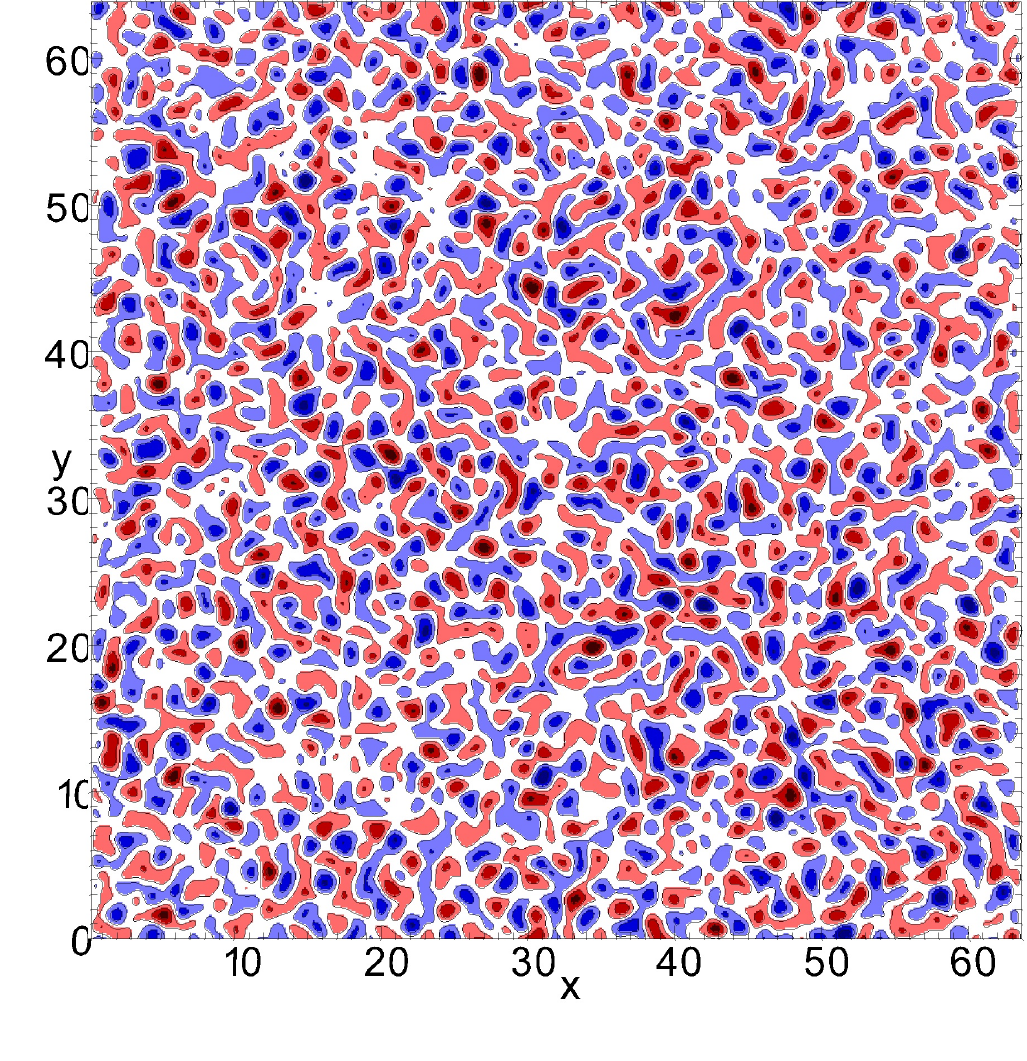}
    \includegraphics[trim = 60px 310px 135px 101px, clip, scale=0.32]{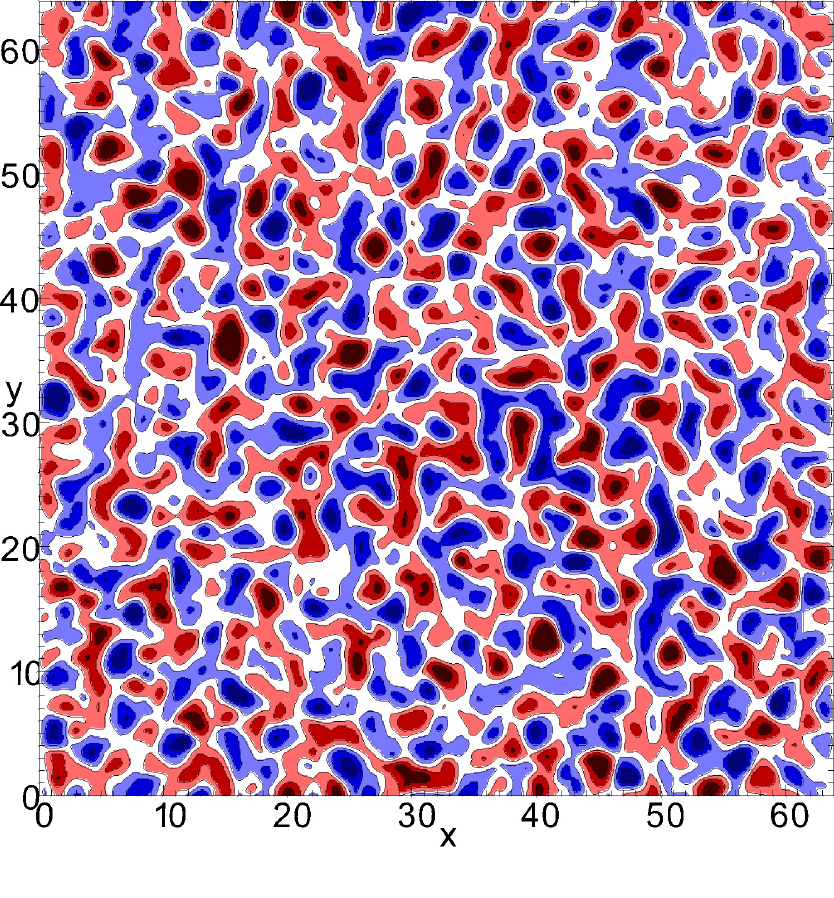}
    \includegraphics[trim = 50px 260px 72px 73px, clip, scale=0.27]{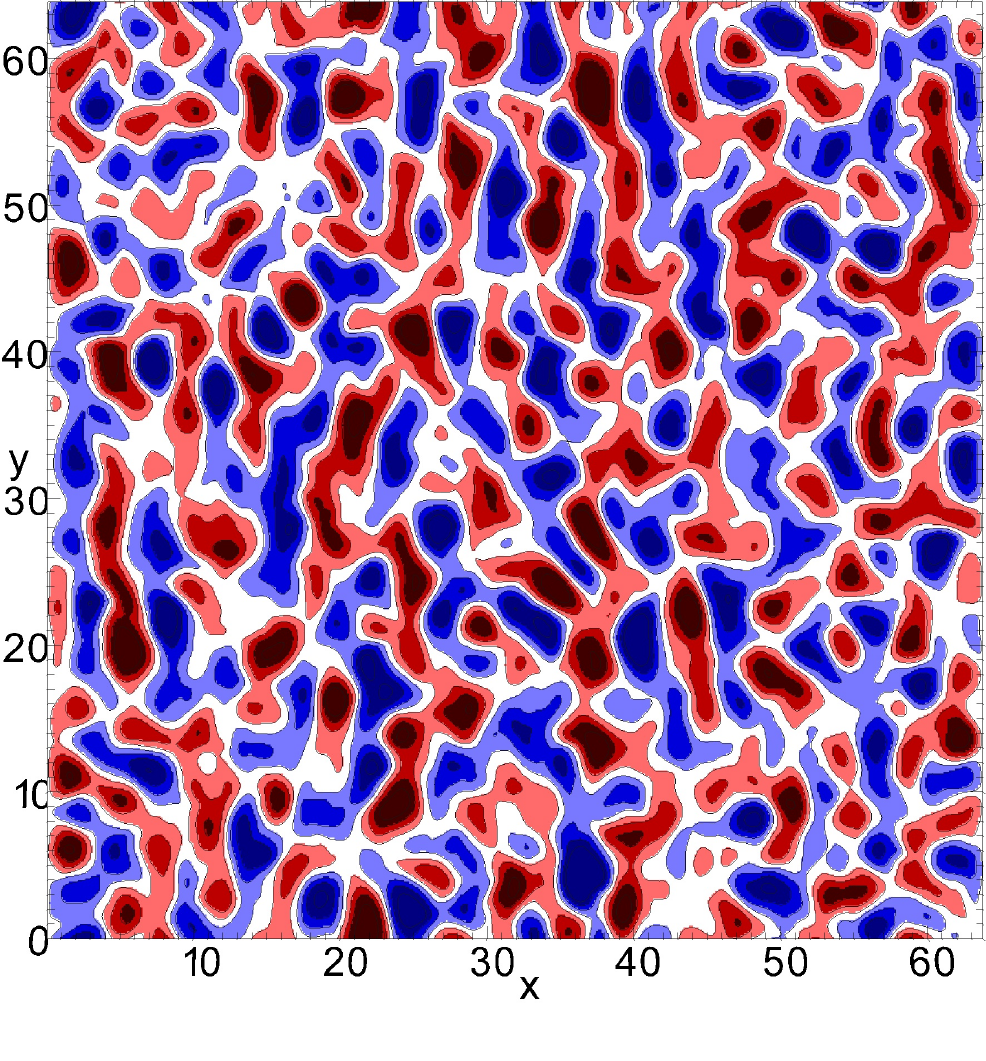}
     \includegraphics[trim = 250px 380px 290px 160px,clip,scale=0.45]{colorlegend_0_5.pdf}
\caption{\label{fig:decay} This electrostatic potential $\delta \phi$ is
  shown at several times for the FD scheme (first row) and the LBM (second row). The tendency to large scale zonally extended structures is visible
  in the advanced state ($t=90$) of the decaying turbulence.} 
\end{figure*}

The initial energy spectrum ${E}_{\delta \phi}({k}) \propto
{k}^{30} ({k}+{k}_0)^{-60}$ is determined in k-space by a random
phase factor and a narrow peak around a wavenumber ${k}_0$ \cite{iga03,larichev91},
which is here is fixed to ${k}_0 = 0.5$. 
Hence around 32 field modes are initialized into a physical box size of $64 \rho_s$. 
The initial amplitude factor of the electrostatic potential field is chosen
that both algorithms remain stable during the computation time, whereby the LBM
is more restrictive, and $\mathcal{O} (\delta{\phi}) = 0.5$ is used as
an examplary value. This sets the initial values of the total generalized
energy and  total generalized enstrophy to $\mathcal{E} \approx 0.07$
and $\mathcal{U} \approx 0.7$, which are defined by   
\begin{eqnarray}
 \mathcal{{E}} &=& \frac{1}{N_x N_y} \sum_{i=0}^{N_x-1}\sum_{j=0}^{N_y-1}
 \frac{1}{2}\left[ \delta \phi^2 + (\vec{\nabla} \delta
   \phi)^2  \right], \\ 
 \mathcal{U} &=& \frac{1}{N_x N_y}  \sum_{i=0}^{N_x-1}\sum_{j=0}^{N_y-1}
 \frac{1}{2}\left[ (\vec{\nabla} \delta \phi)^2 +
   (\vec{\nabla}^2 \delta \phi)^2  \right].
\end{eqnarray}
The spatial gradients are related to the macroscopic velocity via the lowest order momentum balance equation 
\(\vec{e}_z \times \vec{u} = -\vec{\nabla}\delta\phi\). Hence the kinetic energy and the enstrophy
are derived by \( (\vec{\nabla} \delta \phi)^2 =   \vec{u}^2 \) and \( (\vec{\nabla}^2 \delta \phi)^2 = (\partial_x u_y - \partial_y u_x)^2\).
The partial derivatives are computed with the help of central differences of second order accuracy.
Fig.~\ref{fig:decay} shows the decaying turbulent potential field. Up
to $t\approx 30$ the turbulent field is nearly isotropic.
The emerging anisotropy, visible through a pattern of elongated structures
into y-direction, gradually increases as the simulation advances.
The one-dimensional generalized energy spectra ${E} ({k}_x)$ and
${E}({k}_y)$ differ by a few orders of magnitude in the high
${k}_x$, ${k}_y$ range \cite{manfredi99}, apart from the dumb-bell shaped 
two-dimensional generalized energy spectrum (cf.~\cite{naulin02}). 
The correlation between the turbulent electrostatic potential fields of the LB and FD scheme
decreases as time progresses. 

%................

Fig.~\ref{fig:kshellgenenstrgenenerg} shows the corresponding angle-averaged generalized
energy spectrum ${E}({k})$ of the nearly isotropic state which is
defined as the sum over the energy shell within ${k} \pm \Delta {k}$: 
\begin{eqnarray}
 {E}({k}) &=&
 \sum_{\vec{{k}}} {E}(\vec{{k}}) , \hspace{5mm} {k} -\Delta
 {k}< |\vec{{k}}| <{k} +\Delta {k} \\ 
{E}(\vec{{k}})&=& \frac{1}{2} \left( | \delta
\phi_{\vec{{k}}} |^2 + |\vec{{k}} \; \delta
\phi_{\vec{{k}}} | ^2\right) 
\end{eqnarray}
Due to the non-periodicity of the signal in $x$-direction a Blackmann-Harris
window \cite{harris78} has been used on the Fourier transform. 
As a result the transformed signals do not alter the overall power law
coefficient of the $k$-spectrum. The obtained power law coefficients resemble
the theoretically \cite{ottaviani92} and numerically
\cite{iga03,fontan95,kukharkin95} predicted strong turbulence laws   
\begin{equation} {E}({k}) \propto
\begin{cases}
 {k}^4 &,{k} < {k}_{max} \\ 
 {k}^{-5}&, {k}_{max} < {k} \ll 1 \\ 
 {k}^{-3}&, 1 \ll {k}
\end{cases}
\end{equation}
except in the high-$k$ dissipative range, where the power law coefficient
steepens to approximately ${k}^{-4}$. 
The LBM spectrum moreover reveals a weak peak in the very
high $k$ range arising from the residual force term contributions of the
free-slip boundary condition. 

The time evolution of the generalized energy $\mathcal{{E}}$ and
enstrophy $\mathcal{U}$ for both algorithms are shown in
In Fig.~\ref{fig:kshellgenenstrgenenerg}
the time evolution of the generalized energy $\mathcal{{E}}$ and
enstrophy $\mathcal{U}$ for both algorithms resembles decay power laws.
Due to the different treatment of the viscous dissipation the corresponding
decay is slower for the hyperviscvous implementation in the FD scheme.  
The deviation at $t=0$ of the initial variables is based in the
differing resolutions underlying the Fourier transforms of the $k$-spectra, 
and in the initialization of the dynamical variables itself. 
The fitted power law coefficients are of same magnitude as the estimates of $\mathcal{U} \propto
t^{-0.5}$ and $\mathcal{{E}} \propto t^{-0.05}$ \cite{iga03}. 

%................
\begin{figure*}[!ht]
  \centering
  \caption*{(a)\hspace{50 mm}(b)\hspace{50 mm}(c)}
  \includegraphics[trim = 5px 20px 0px 20px,  clip, scale=0.40]{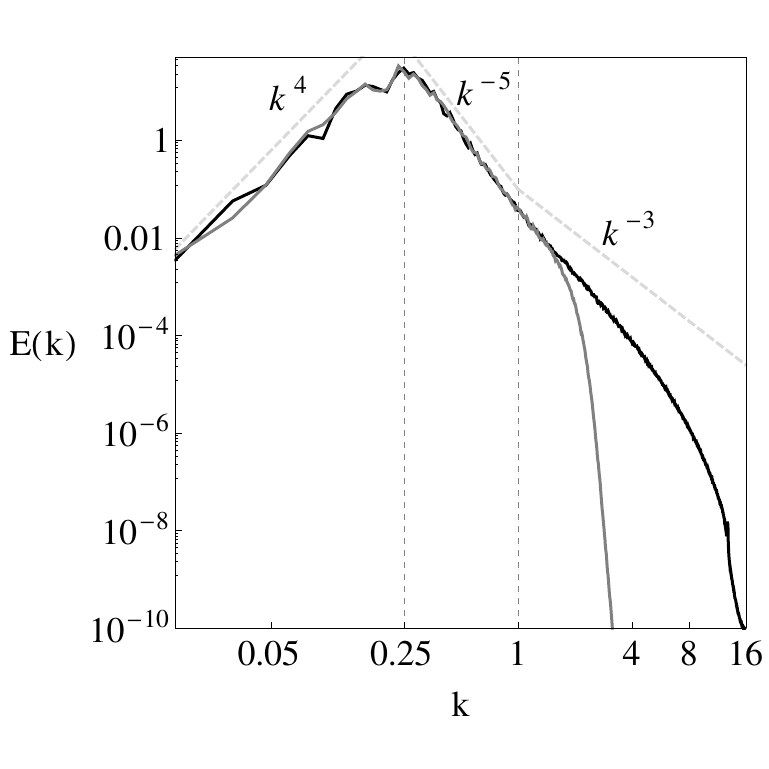}
  \includegraphics[trim = -5px 0px 0px 15px,  clip, scale=0.37]{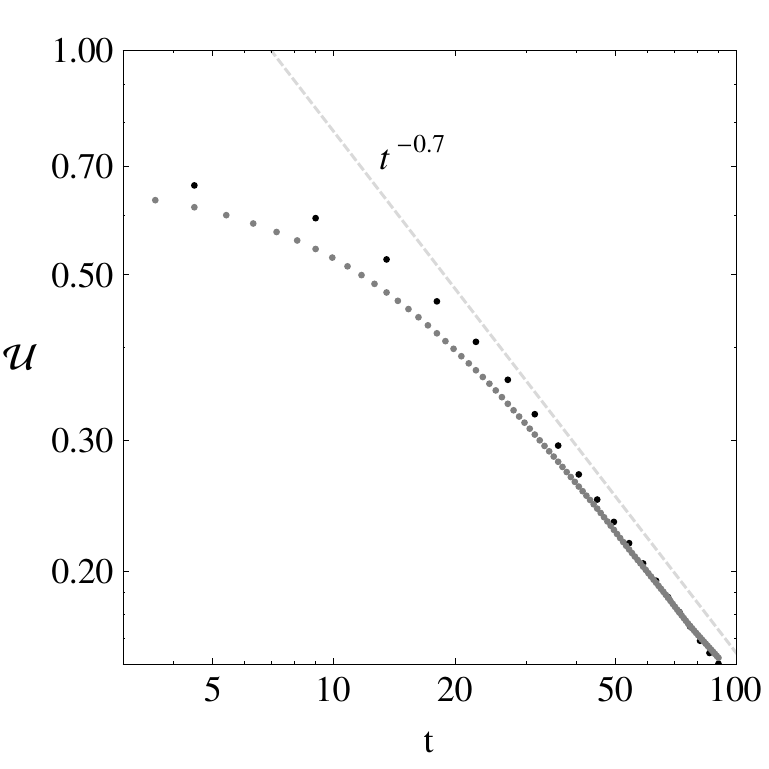}
  \includegraphics[trim = -5px 7px 0px 25px, clip, scale=0.39]{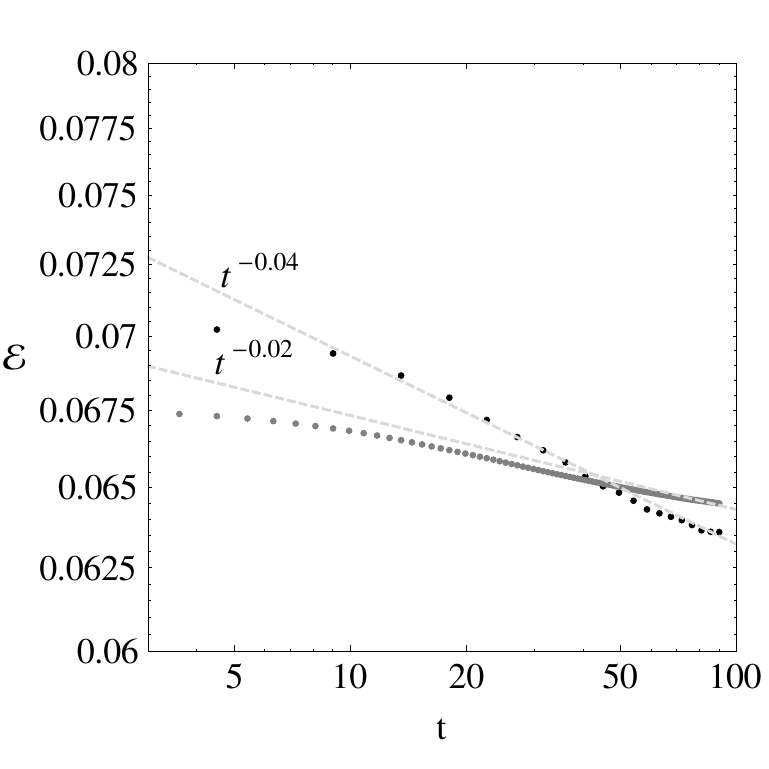}
  \caption{\label{fig:kshellgenenstrgenenerg.pdf}
  (a) Double logarithmic plot of the time
  averaged k-spectrum  of the generalized energy ${E}({k})$. Black line: LBM; gray line:
  FD. Time averaging was applied between $t=9$ and $t=32$.
  (b) Double logarithmic plot of time evolution of the
  generalized enstrophy $\mathcal{U} ( t)$ for the LBM (black dots) and the FD scheme (gray dots).
  (c) Double logarithmic plot of time evolution of the
  generalized energy $\mathcal{{E}} ( t)$ for the LBM (black dots) and the FD scheme (gray dots). }
\end{figure*}

%................
% \begin{figure*}[h]
%  \includegraphics[trim = -35px 0px 0px 0px, clip, clip, scale=0.45]{genenstr_corr}
% \caption{\label{fig:genenstr} Double logarithmic plot of time evolution of the
%   generalized enstrophy $\mathcal{U} ( t)$ for the LBM (black dots) and the FD scheme (gray dots). }
% \end{figure*}
% 
% %................
% \begin{figure*}[h]
%  \includegraphics[trim = -35px 0px 0px 0px, clip, clip, scale=0.45]{genenerg_corr}
% \caption{\label{fig:genenerg} Double logarithmic plot of time evolution of the
%   generalized energy $\mathcal{{E}} ( t)$ for the LBM (black dots) and the FD scheme (gray dots).  }
% \end{figure*}

\section{\label{sec:conc}Conclusion}

The presented LBM algorithm for the CHM equation is based on previous
single-layer shallow water LBM implementations with an additional source term,
which is able to include density gradient effects. 
Consequently the form of the forcing function is revised to reproduce the
correct macroscopic equations in the course of the asymptotic analysis. 
In order to verify the scheme, computations of decaying turbulence, dipole
drift modons and monopole propagation with the new LBM scheme were
compared with an established FD scheme. 

The numerical results deviate mainly in the observed (in)stability of the
drift modon from that of the CHM equation, 
and resemble apart from that characteristic drift wave turbulence
behaviour. 
The occuring shear in the $x$-direction reduces with the drift parameter \(\kappa_n\) and
persists even if the approximated velocityshifts (eqs. (\ref{approxzerothvelocitymoment})
and (\ref{approxfirstvelocitymoment})) are replaced by the exact expressions. 
Hence the shear effect is intrinsically related to the compressible limit of the 
resolved ion continuity and momentum equations.

Alternatively, there is another option to approach the CHM equation via a LB model
by replacing the density gradient source term, which appears in the continuity
equation with a spatially varying gyro frequency (or Coriolis parameter), by  
a substitution of $\left( \vec{e}_z \times \vec{u}\right) \rightarrow
\left(1+\kappa_n {x}\right) \left(\vec{e}_z \times \vec{u} \right)$
in the momentum equation. 
As a result the implicit equations comparable with  
eqs.~(\ref{approxzerothvelocitymoment}) and (\ref{approxfirstvelocitymoment})
yield simple expressions without any further
approximations \cite{dellar01}. However, correponding computations for $\kappa_n = 0.05$
showed a considerable larger shear in the $x$-direction as the presented LBM
algorithm. 

%............................................................................................

 \section*{Acknowledgement}
This work was partly supported by the Austrian Science Fund (FWF) Y398; 
by the Austrian Ministry of Science BMWF as part of the UniInfrastrukturprogramm
of the Research Platform Scientific Computing at the University of Innsbruck;
and by the European Commission under the Contract of Association between
EURATOM and \"OAW carried out within the framework of the European Fusion
Development Agreement (EFDA). The views and opinions expressed herein do not
necessarily reflect those of the European Commission.

%............................................................................................
\appendix

\section{\label{app:aa}Asymptotic analysis}
By adopting the diffusive scaling \(\delta t \sim \delta x^2 \sim \epsilon^2 \) in our LB approach, an asymptotic expansion
of the model equations in the spirit of Sone \cite{sone02}, Junk \textit{et al.} \cite{junk05,junk03} and Inamuro \textit{et al.} \cite{inamuro97}, will lead directly to an incompressible set of fluid type equations by
taking the Mach number and Knudsen number to zero concurrently, while fixing the Reynolds number. Compressibility effects 
are then related to the diffusive time scale and are understood as numerical artifacts instead of physical effects.
In contrast to the Chapman-Enskog (CE) derivation the ordering of the macroscopic variables is not made beforehand, so that the expansion
of the macroscopic variables is ambiguous and no further limiting process (e.g. low Mach number) has to be applied to obtain the incompressible fluid equations. 
To clearly demonstrate the deviations from the incompressible fluid equations up to a specific order in \(\epsilon\), 
we start the derivation from the scaled difference equation (LBE). This is more accurate
then just analyzing the discrete Boltzmann PDE and additionally it will alter some terms of the underlying incompressible fluid type equations. We 
start our analysis from the explicit LB eq. (\ref{explicitscaledLBE}), where we 
expand the distribution function and the forcing function in an asymptotic series of \(\epsilon\) according to
\begin{align}
  \bar{f}_{i} ({\vec{x}},t)  &= \sum_{m=0}^{\infty}  \epsilon^{m}  \bar{f}_{i}^{(m)}  ({\vec{x}},t),  &
    F_{i} ({\vec{x}},t)  &= \sum_{m=0}^{\infty}  \epsilon^{m}  F_{i}^{(m)} ({\vec{x}},t),
\end{align}
whereby the leading order appears at \(\mathcal{O}(\epsilon^3)\) for the forcing function and naturally \(F_{i}^{(0)}= F_{i}^{(1)}=F_{i}^{(2)}=0\).
A Taylor approximation of the left-hand side of the LB eq. (\ref{explicitscaledLBE}) provides the quantitiy
\begin{align}
  \bar f_{i}^{(m)}({\vec{x}}',t') -  \bar f_{i}^{(m)} ({\vec{x}},t) &=  \sum_{r=0}^{\infty} \epsilon^r D_r(\partial_t,\vec{\xi}_i \cdot \vec{\nabla} ) \bar f_{i}^{(m)} ({\vec{x}},t) 
 \end{align}
with the polynomials given in general by
\begin{align}
 D_r (\partial_t,\vec{\xi}_i \cdot \vec{\nabla}) &= \sum_{2a+b=r} \frac{(\partial_t)^a(\vec{\xi}_i \cdot \vec{\nabla})^b}{a! b!},& r \geq 0
\end{align}
and practically by
\begin{align}
 D_0(\partial_t,\vec{\xi}_i \cdot \vec{\nabla} ) &= 0, \\ 
 D_1(\partial_t,\vec{\xi}_i \cdot \vec{\nabla} ) &=\left( \vec{\xi}_i \cdot \vec{\nabla} \right), \\
 D_2(\partial_t,\vec{\xi}_i \cdot \vec{\nabla} ) &=\partial_t +\left( \vec{\xi}_i \cdot \vec{\nabla} \right)^2/2, \\
 D_3(\partial_t,\vec{\xi}_i \cdot \vec{\nabla} ) &=\left( \vec{\xi}_i \cdot \vec{\nabla} \right) \left[\partial_t
 +\left( \vec{\xi}_i \cdot \vec{\nabla} \right)^2/6  \right].
\end{align}
For the sake of convenience the equilibrium distribution function is split into three parts
\(f_i^{(eq)} =f_i^{(eq,0)}+ f_i^{(eq,1)} + f_i^{(eq,2)}\) by analogy with Asinari \cite{asinari08}, whereby the nonvanishing first three moments over the equilibrium distribution functions are given by
\begin{align}
  \sum\limits_{i=0}^8 f_i^{(eq,0)}(\phi^{(k)}) &= \phi^{(k)}, \\ 
  \sum\limits_{i=0}^8 \vec{\xi}_i \vec{\xi}_if_i^{(eq,0)}(\phi^{(k)}) &= P^{(k)} \vec{I}, \\
  \sum\limits_{i=0}^8 \vec{\xi}_i f_i^{(eq,1)} (\phi^{(q)},\vec{u}^{(p)})&= \phi^{(q)} \vec{u}^{(p)}, \\
  \sum\limits_{i=0}^8 \vec{\xi}_i \vec{\xi}_i f_i^{(eq,2)}(\phi^{(q)},\vec{u}^{(p)},\vec{u}^{(r)}) &= \phi^{(q)} \vec{u}^{(p)} \vec{u} ^{(r)}.
\end{align}
Combining now the expansions for \(\bar{f}_i\), \(f_i^{(eq)}\) and \(F_i\) with eq. (\ref{explicitscaledLBE})
yields a discrete PDE of order \(\mathcal{O}(\epsilon^{k+2})\) with \( k\geq 2\)
\begin{align}
\label{discretePDEscaled}
 \partial_t&  \bar f_{i}^{(k)}  + \left(\vec{\xi}_i \cdot \vec{\nabla}  \right) \bar f_{i}^{(k+1)} +
 \frac{1}{2} \left(\vec{\xi}_i \cdot \vec{\nabla}  \right)^2 \bar f_{i}^{(k)} =  \nonumber \\& 
 \bar A  \Bigg\{
 f_i^{(eq,0)}(\phi^{(k+2)}) + 
  \sum_{p+q=k+2} f_i^{(eq,1)}(\phi^{(q)},\vec{u}^{(p)}) +\Bigg.  \nonumber\\  & \Bigg. 
  \sum_{p+q+r=k+2} f_i^{(eq,2)}(\phi^{(q)},\vec{u}^{(p)},\vec{u}^{(r)})  - \bar f_i^{(k+2)} )  \Bigg\}+ L^{(k+2)}
\end{align}
and the general definition 
\begin{align}
 L_i^{(k+2)} &= \bar \lambda F_i^{(k+2)} -\sum_{\substack{m+r=k+2{}\\m < k}} D_r(\partial_t,\vec{\xi}_i \cdot \vec{\nabla} )  \bar  f_{i}^{(m)} ,   
 \end{align}
 revealing differences to an asymptotic analysis of the discrete Boltzmann eq. at \(\mathcal{O}(\epsilon^3)\):
 \begin{align}
 L_i^{(0)} &= L_i^{(1)} =L_i^{(2)}= 0, \\
 L_i^{(3)} &= \bar \lambda  F_i^{(3)} -\left( \vec{\xi}_i \cdot \vec{\nabla} \right) \left[\partial_t
 +\left( \vec{\xi}_i \cdot \vec{\nabla} \right)^2/6 \right]  \bar f_{i}^{(0)}.
\end{align}
Rearranging eq. (\ref{discretePDEscaled}) to
\begin{widetext}
\begin{align}
\label{rearrangeddiscretePDEscaled}
  \bar f_i^{(k+2)} =& f_i^{(eq,0)}(\phi^{(k+2)}) + 
  \sum_{p+q=k+2} f_i^{(eq,1)}(\phi^{(q)},\vec{u}^{(p)}) + 
  \sum_{p+q+r=k+2} f_i^{(eq,2)}(\phi^{(q)},\vec{u}^{(p)},\vec{u}^{(r)}) \\ \nonumber &
  -\bar A^{-1} \left[\partial_t  \bar f_{i}^{(k)}  + \left(\vec{\xi}_i \cdot \vec{\nabla}  \right) \bar f_{i}^{(k+1)} +
 \frac{1}{2} \left(\vec{\xi}_i \cdot \vec{\nabla}  \right)^2 \bar f_{i}^{(k)} \right]  +\bar  A^{-1} L^{(k+2)}
\end{align}
allows us to construct the expansion coefficients \( f^{(k)} \) by induction of eq. (\ref{rearrangeddiscretePDEscaled}). The first three reduce to
\begin{align}
\bar f_i^{(0)} &= f_i^{(eq,0)} (\phi^{(0)}),\\
 \bar f_i^{(1)} &= f_i^{(eq,0)} (\phi^{(1)}) + f_i^{(eq,1)} (\phi^{(0)},\vec{u}^{(1)})  -\bar A^{-1}  \left(\vec{\xi}_i \cdot \vec{\nabla}  \right) \bar f^{(0)},\\
 \bar f_i^{(2)} &= f_i^{(eq,0)} (\phi^{(2)}) + f_i^{(eq,1)} (\phi^{(0)},\vec{u}^{(2)})+  f_i^{(eq,1)} (\phi^{(1)},\vec{u}^{(1)})
 + f_i^{(eq,2)} (\phi^{(0)},\vec{u}^{(1)},\vec{u}^{(1)}) \nonumber \\  &
 -\bar A^{-1}\left[  \left(\vec{\xi}_i \cdot \vec{\nabla}  \right) \bar f^{(1)} + \left(\partial_t 
 + \frac{1}{2}  \left(\vec{\xi}_i \cdot \vec{\nabla}  \right)^2  \right)\bar f^{(0)}\right].
\end{align}
Introducing now the j-th moment over \(\bar f^{(k)}\), \(L^{(k)}\) and \(F^{(k)}\) by \(\mathcal{M}^{(k)}_j\),\(\mathcal{L}^{(k)}_j\) and \(\mathcal{F}^{(k)}_j\)
and writing down the relevant contributions in more detail 
\begin{align}
  \mathcal{M}^{(0)}_0 &= \phi^{(0)},     &   \mathcal{M}^{(1)}_0 &= \phi^{(1)},      &
  \mathcal{M}^{(2)}_0 &=  \phi^{(2)}  -  \bar A^{-1} \left[\vec{\nabla}  \cdot  ( \phi^{(0)}  \vec{u}^{(1)}   ) 
                                                                 + \vec{\nabla}^2 P^{(0)} /2
                                                                 + \partial_t \phi^{(0)}\right], \\
  \vec{\mathcal{M}}^{(0)}_1      &= \phi^{(0)}  \vec{u}^{(0)}=0 , & \vec{\mathcal{M}}^{(1)}_1      &= \phi^{(0)}  \vec{u}^{(1)},    &
  \vec{\mathcal{M}}^{(2)}_1 &= \phi^{(1)}\vec{u}^{(1)} +\phi^{(0)}\vec{u}^{(2)} -  \bar A^{-1}  \left[ \vec{\nabla}    P^{(1)} \right],\\
  \vec{\mathcal{M}}^{(0)}_2   &=  P^{(0)} \vec{I},  &  
  \vec{\mathcal{M}}^{(1)}_2   &=  P^{(1)} \vec{I}, &  
  \vec{\mathcal{M}}^{(2)}_2 &=P^{(2)}  \vec{I} +  \phi^{(0)} \vec{u}^{(1)}\vec{u}^{(1)} - \bar A^{-1} \mathcal{D}^{(1)}
\end{align}
and
\begin{align}
   \mathcal{F}^{(3)}_0  &= \phi^{(0)} s^{(1)}&
  \vec{\mathcal{F}}^{(3)}_1  &= \phi^{(0)} \vec{a}^{(1)} &
  \vec{\mathcal{F}}^{(3)}_2  &=   \frac{d P^{(0)}}{d \phi^{(0)}} \phi^{(0)} s^{(1)} \vec{I} \\
     \mathcal{L}^{(3)}_0  &=  \bar{\lambda} \mathcal{F}^{(3)}_0  &
  \vec{\mathcal{L}}^{(3)}_1  &=\bar{\lambda} \mathcal{F}^{(3)}_1- \mathcal{C}^{(3)}_1   &
  \vec{\mathcal{L}}^{(3)}_2  &= \bar{\lambda} \mathcal{F}^{(3)}_2 
\end{align}
with the barotropic pressure terms up to \(\mathcal{O}(\epsilon^2)\)
\begin{align}   
 P^{(0)} &= \frac{1}{2 \kappa_n^2}         (\phi^{(0)})^2 &
 P^{(1)} &= \frac{1}{\kappa_n^2}           \phi^{(0)} \phi^{(1)} &
 P^{(2)} &= \frac{1}{2 \kappa_n^2} \left(2 \phi^{(0)} \phi^{(2)} + (\phi^{(1)})^2 \right) 
\end{align}
and the dissipative and residual term for \(k=1\)
\begin{align}
 \vec{\nabla} \cdot \vec{\mathcal{D}}^{(1)}&=\sum\limits_{i=0}^8  \vec{\xi}_i \left(\vec{\xi}_i \cdot \vec{\nabla}\right)^2 f^{(1)}  = 
\theta \left[\vec{\nabla}^2 \left(\phi^{(0)} \vec{u}^{(1)}\right) + 2 \vec{\nabla} \left(\vec{\nabla}\cdot \left(\phi^{(0)} \vec{u}^{(1)}\right) \right)\right] \\
\vec{\mathcal{C}}^{(3)}_1&=\sum\limits_{i=0}^8  \vec{\xi}_i    \left( \vec{\xi}_i \cdot \vec{\nabla} \right) \left[\partial_t
 +\left( \vec{\xi}_i \cdot \vec{\nabla} \right)^2/6 \right]  \bar f_{i}^{(0)}
\end{align}
Taking the zeroth and first moment over eq. (\ref{discretePDEscaled}) yields 
\begin{align}
\label{conteq}
 \partial_t \mathcal{M}^{(k)}_0 + \vec{\nabla} \cdot \vec{\mathcal{M}}^{(k+1)}_1 
 +\frac{1}{2}  \vec{\nabla} \vec{\nabla} : \vec{\mathcal{M}}^{(k)}_2&= \mathcal{L}^{(k+2)}_0 \\
 \label{momeq}
 \partial_t   \vec{\mathcal{M}}^{(k)}_1  + \vec{\nabla} \cdot \vec{\mathcal{M}}^{(k+1)}_2
 +\frac{1}{2}\vec{\nabla} \cdot  \vec{\mathcal{D}}^{(k)}  &=  \vec{\mathcal{L}}^{(k+2)}_1 
\end{align}
For \(k=-1\) eq. (\ref{momeq}) delivers following constraint for the pressure \( \vec{\nabla}  P^{(0)}  = 0 \), allowing us to choose
\(\phi^{(0)}=1\) due to \(\phi^{(0)}\vec{\nabla} \phi^{(0)} =0\) resulting in \(\vec{u}^{(0)}=0\).  For \(k=0\) eq. (\ref{conteq}) reduces  to
\begin{align}
\label{conteq0}
 \partial_t \phi^{(0)} + \vec{\nabla} \cdot  \phi^{(0)}  \vec{u}^{(1)} &= 0 
\end{align}
whereas eq. (\ref{momeq}) yields \(\vec{\nabla}  P^{(1)}  = 0\), which permits us to choose \(\phi^{(1)}=0\) due to \(\phi^{(1)}\vec{\nabla} \phi^{(0)} +\phi^{(0)}\vec{\nabla} \phi^{(1)}=0\). As a consequence
the pressure terms up to \(\mathcal{O}(\epsilon^3)\) are fixed to \(  P^{(0)}=1/(\kappa_n^2)\), \(  P^{(1)}=0\), \(  P^{(2)}=1/(\kappa_n^2)\phi^{(2)}\) and \(  P^{(3)}=1/(\kappa_n^2)\phi^{(3)}\).
From eq. (\ref{conteq0}) the incompressibility condition for \( \vec{u}^{(1)} \) follows immediately
\begin{align}
\label{conteq0inc}
 \vec{\nabla} \cdot   \vec{u}^{(1)} &= 0 .
\end{align}
Applying these properties for eqs. (\ref{conteq}) and (\ref{momeq}) for \(k=1\) results in
\begin{align}
\label{conteq1}
 \partial_t\phi^{(1)}+ \vec{\nabla} \cdot \left[  \phi^{(1)}\vec{u}^{(1)} +\phi^{(0)}\vec{u}^{(2)} \right]
&= \bar{\lambda} \phi^{(0)} s^{(1)}, \\
 \label{momeq1}
 \partial_t  \phi^{(0)}  \vec{u}^{(1)} + \vec{\nabla} \cdot \left[\phi^{(0)} \vec{u}^{(1)}\vec{u}^{(1)} +P^{(2)}  \vec{I} \right]
 +\left(\frac{1}{2} 
  - \frac{1}{\bar A}\right) \vec{\nabla} \cdot \vec{\mathcal{D}}^{(1)} &=\bar{\lambda}\phi^{(0)}\vec{a}^{(1)} - \vec{\mathcal{C}}^{(3)}_1 
\end{align}
Substituting the constraints for \(\phi^{(0)}=1\) and \(\phi^{(1)}=0\) into the latter two equations we obtain compressibility effects
at \(\mathcal{O}(\epsilon^2)\) due to the intrinsic source term. Additionally the residual term on the right hand side of eq. \(\ref{momeq1}\) vanishes
\begin{align}
\label{conteq1inc}
  \vec{\nabla} \cdot \vec{u}^{(2)}
&= \bar{\lambda}  s^{(1)},\\
 \label{momeq1inc}
  \partial_t    \vec{u}^{(1)} + \vec{\nabla} \cdot  \left(\vec{u}^{(1)}\vec{u}^{(1)}\right) +\frac{1}{\kappa_n^2}  \vec{\nabla} \phi^{(2)} 
 +\theta \left(\frac{1}{2} 
  - \frac{1}{\bar A}\right) \vec{\nabla}^2   \vec{u}^{(1)} &=\bar{\lambda}\vec{a}^{(1)} .
\end{align}
For \(k=2\) for eqs. (\ref{conteq}) and (\ref{momeq}) the derivation gives
\begin{align}
\label{conteq2}
 \partial_t\phi^{(2)}+ \vec{\nabla} \cdot \left[  \phi^{(2)}\vec{u}^{(1)} +\phi^{(0)}\vec{u}^{(3)} \right]
&= \bar{\lambda} \phi^{(0)} s^{(2)},   \\
 \label{momeq2}
 \partial_t  \phi^{(0)}  \vec{u}^{(2)} +  \vec{\nabla} \cdot \left[\phi^{(0)} \left(\vec{u}^{(1)}\vec{u}^{(2)}+\vec{u}^{(2)}\vec{u}^{(1)} \right)+P^{(3)}  \vec{I} \right]
 +\left(\frac{1}{2} 
  - \frac{1}{\bar A}\right) \vec{\nabla} \cdot \vec{\mathcal{D}}^{(2)}&=\bar{\lambda}\phi^{(0)}\vec{a}^{(2)} - \vec{\mathcal{C}}^{(4)}_1  ,
\end{align}
which reduces analogeously to
\begin{align}
\label{conteq2inc}
 \partial_t\phi^{(2)} + \vec{u}^{(1)}\cdot \vec{\nabla}\phi^{(2)} +\vec{\nabla} \cdot \vec{u}^{(3)}
&= \bar{\lambda}  s^{(2)}, \\
 \label{momeq2inc}
\partial_t    \vec{u}^{(2)} +  \vec{\nabla} \cdot  \left(\vec{u}^{(1)}\vec{u}^{(2)}+\vec{u}^{(2)}\vec{u}^{(1)} \right) +\frac{1}{\kappa_n^2}  \vec{\nabla} \phi^{(3)} 
 +\theta \left(\frac{1}{2} 
  - \frac{1}{\bar A}\right)\left[ \vec{\nabla}^2   \vec{u}^{(2)} + 2 \vec{\nabla} \left(\vec{\nabla} \cdot \vec{u}^{(2)} \right) \right] &=\bar{\lambda}\vec{a}^{(2)} .
\end{align}
To show the deviations from the anticipated model equations (\ref{ioncont}) and (\ref{ionmom}) at  \(\mathcal{O}(\epsilon^3)\) we multiply
eq. (\ref{conteq0inc}) with \(\epsilon\),
eq. (\ref{conteq1inc}) with \(\epsilon^2\),
eq. (\ref{conteq2inc}) with \(\epsilon^3\), 
eq. (\ref{momeq1inc})  with \(\epsilon^2\) and
eq. (\ref{momeq2inc})  with \(\epsilon^3\). 
Summing them up and introducing  quantities up to order \(\mathcal{O}(\epsilon^3)\)
\begin{align}
 \tilde{\phi} &= 1 + \epsilon^2  \phi^{(2)}, & \tilde{\vec{u}} &= \epsilon \vec{u}^{(1)}+\epsilon^2 \vec{u}^{(2)} &\tilde{P} &= \epsilon^2 P^{(2)}+ \epsilon^3 P^{(3)} ,
 \end{align}
 \begin{align}
  \tilde{s} &= \epsilon s^{(1)}+\epsilon^2 s^{(2)} = \kappa_n \left(\epsilon u_x^{(1)} + \epsilon^2 u_x^{(2)} \right), &
 \tilde{\vec{a}} &= \epsilon \vec{a}^{(1)}+ \epsilon^2 \vec{a}^{(2)} =
  \kappa_n^{-1}  \left(\epsilon \vec{u}^{(1)} + \epsilon^2 \vec{u}^{(2)}\right)\times \vec{e}_z   +
 \epsilon^2 \vec{u}^{(1)}  s^{(1)} 
\end{align}
yields with the viscosity modification \( \tilde{\nu} = \theta \left(1/A - 1/2 \right)  \) and the operator 
\(d^{(1)}_t= \partial_t  +  \vec{{u}}^{(1)}  \cdot \vec{\nabla} \)
\begin{align}
\label{conteq01}
 \epsilon \frac{d^{(1)}}{d t} \tilde \phi + \epsilon^3\vec{\nabla} \cdot \vec{u}^{(3)} + \vec{\nabla} \cdot \vec{\tilde{u}}
&= \epsilon \bar{\lambda} \kappa_n  \tilde{u}_x  + \mathcal{O}(\epsilon^4),\\
 \label{momeq01}
\epsilon \frac{d^{(1)}}{d t}\vec{\tilde{u}}
+ \epsilon^3 \vec{{u}}^{(2)}  \cdot \vec{\nabla} \vec{u}^{(1)}    
 +\epsilon \bar{\lambda} \kappa_n^{-1} \vec{e}_z \times \tilde{\vec{u}}  &=- \vec{\nabla} \tilde{P} +  \epsilon \tilde{\nu} \left[\vec{\nabla}^2 \vec{\tilde{u}}+ 2 \vec{\nabla} \left(\vec{\nabla} \cdot \epsilon^2 \vec{u}^{(2)}\right)\right]
 +\mathcal{O}(\epsilon^4).
\end{align}
\end{widetext}
In eq. (\ref{momeq01}) the source term appearing in \( \vec{a}^{(2)} \) cancels an additional term arising from the advective derivative term at \(\mathcal{O}(\epsilon^3)\).
The result shows that the approximation to the magnetised plasma equations (\ref{ioncont}) and (\ref{ionmom}) are at least second order accurate and that the deviations from this set appear at \(\mathcal{O}(\epsilon^3)\). As a consequence
of the diffusive scaling this refers to second-order accuracy in space and first-order accuracy in time. Moreover  
the Newtonian deviatoric stress  ${\vec{\sigma}}'_{N} =  \tilde {\nu}  \tilde \phi [ ( {\vec\nabla}
  {\vec{ \tilde u}} ) +  ( {\vec\nabla} {\vec{  \tilde u}} )^T - \frac{1}{2}
  \left({\vec\nabla} \cdot   {\vec{ \tilde u}}\right)  \vec{I} ] + \zeta  \tilde \phi \left({\vec\nabla} \cdot   {\vec{ \tilde u}}\right)  \vec{I}  $ appears
  with an artificial bulk viscosity $\zeta = (5/3) \tilde \nu$  at \(\mathcal{O}(\epsilon^4)\) \cite{dellar02pre}.

%..................................................................
%\bibliographystyle{elsearticle-num}

\end{document}